\documentclass[%
 reprint,
superscriptaddress,
 amsmath,amssymb,
 aps,
]{revtex4-2}
\usepackage{color}
\usepackage{gensymb}
\usepackage{graphicx}
\usepackage{dcolumn}
\usepackage{bm}
\usepackage{hyperref}
\usepackage{epstopdf}
\usepackage{float}
\usepackage{listings}
\usepackage{times,mathptm,bm}
\usepackage{xcolor}
\usepackage[export]{adjustbox}

\begin{document}

\title{Locomotion of Active Polymerlike Worms in Porous Media}

\author{R.~Sinaasappel}
\affiliation{Van der Waals-Zeeman Institute, Institute of Physics, University of Amsterdam, 1098XH Amsterdam, The Netherlands.}
\author{M.~Fazelzadeh}
\affiliation{Institute for Theoretical Physics, University of Amsterdam, Science Park 904, 1098XH Amsterdam, The Netherlands.}
\author{T.~Hooijschuur}
\affiliation{Van der Waals-Zeeman Institute, Institute of Physics, University of Amsterdam, 1098XH Amsterdam, The Netherlands.}
\affiliation{Institute for Theoretical Physics, University of Amsterdam, Science Park 904, 1098XH Amsterdam, The Netherlands.}
\author{Q.~Di}
\affiliation{Institute for Theoretical Physics, University of Amsterdam, Science Park 904, 1098XH Amsterdam, The Netherlands.}
\author{S.~Jabbari-Farouji}
\email{s.jabbarifarouji@uva.nl}
\affiliation{Institute for Theoretical Physics, University of Amsterdam, Science Park 904, 1098XH Amsterdam, The Netherlands.}
\author{A.~Deblais}
\email{a.deblais@uva.nl}
\affiliation{Van der Waals-Zeeman Institute, Institute of Physics, University of Amsterdam, 1098XH Amsterdam, The Netherlands.}
\date{\today}

\begin{abstract}
We investigate the locomotion of thin, living \textit{T.~Tubifex} worms, which display active polymerlike behavior, within quasi-2D arrays of cylindrical pillars, examining varying spatial arrangements and densities. These active worms spread in crowded environments, with a dynamics dependent on both  the concentration and arrangement of obstacles. In contrast to passive polymers, our results reveal that in disordered configurations, increasing the pillar density enhances the long-time diffusion of our active polymer-like worms, while we observe the opposite trend in ordered pillar arrays. We found that in disordered media, living worms reptate through available curvilinear tubes, whereas they become trapped within pores of ordered media. Intriguingly, we show that reducing the worm's activity significantly boosts its spread, enabling passive sorting of worms by activity level. Our experimental observations are corroborated through simulations of the tangentially driven polymer model with matched persistence length predicting the same trends.
\end{abstract}

\maketitle

Biological organisms exhibit diverse features to optimize survival and navigate through disordered habitats \cite{dorgan_meandering_2013, Bechinger2016, sosa_motility_2017, makarchuk_enhanced_2019, Bhattacharjee2019, kudrolli_burrowing_2019, dehkharghani_self-transport_2023, Juarez_2010}. Active filaments, from actin filaments in the cytoskeleton to cyanobacteria and earthworms, move through complex, crowded environments. Despite their importance, the mechanisms by which motile filaments navigate porous media remain poorly understood \cite{lux_motility_2001, thornlow_persistent_2015, toley_motility_2012}.

Few, mostly theory-based modeling studies have been performed on the dynamics of active filaments in complex environments \cite{theeyancheri_active_2023,majmudar_experiments_2012,Mokthari2019,Kurzthaler2021,Tejedor2023,Fazelzadeh2023}. These investigations have predominantly focused on active Brownian and tangentially driven polymer models, delving into the interplay of length, flexibility, and activity on polymer dynamics, both within porous media \cite{Mokhtari2019,Kurzthaler2021,Tejedor2023,Fazelzadeh2023} and confined spaces \cite{martinroca2024tangentially}. In the case of active stiff or semi-flexible polymers, they move through the porous medium persistently with motion that closely resembles reptation, a concept originally elucidated by de Gennes for a passive polymer performing wormlike displacements in arrays of fixed obstacles~\cite{deGennes1971}. In the case of very flexible active polymers however, they will bundle up inside the pores, causing hopping-trapping dynamics where the polymers are stuck in the pores and only rarely hop between pores \cite{Mokhtari2019,Fazelzadeh2023}. In these investigations, the effect of medium porosity and particularly the geometry of the pattern on the dynamics of active polymers has received little attention.

\begin{figure}[b]
\centering
\includegraphics[width=0.9\columnwidth]{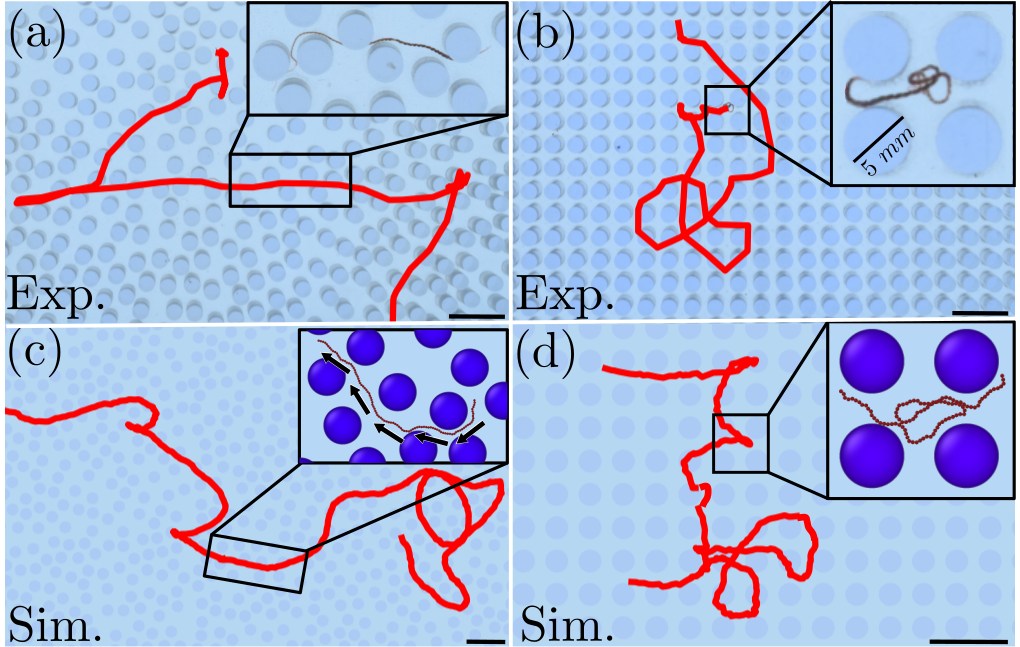}
\caption{Trajectories of the center of mass of an active polymer-like worm in a $L^2$ = 44$\times$44 cm$^2$ 2D pillar array with a 5 mm diameter and surface fraction $\phi$ = 0.4, shown for (a) random packing and (b) square lattice over 185 s (partial view). Close-ups highlight distinct worm conformations in the two obstacle arrangements. (c),(d) Simulated center-of-mass trajectories and polymer conformations of tangentially driven active filaments in the same geometries as in (a),(b). Scale bar is 15 mm. See also Sup.~Vid.~2, 3 \& 5, 6.} 
\label{fig:Trajectories}
\end{figure}

\begin{figure}
\centering
\includegraphics[width=0.9\columnwidth]{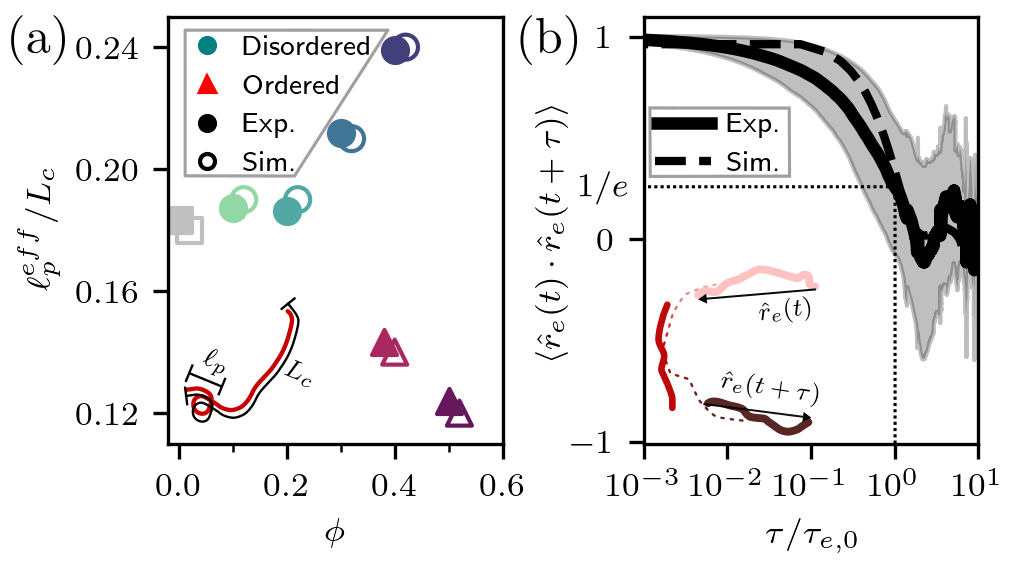}
\caption{(a) Effective persistence length $\ell_p^{eff}/L_c$ of living worms and tangentially driven active polymers as a function of $\phi$. (b) Time auto-correlation function of the end-to-end vector, normalized by the reorientational relaxation time $\tau_{e,0}$ (continuous line, $\tau_{e,0} \sim 40$ s for the worm) and for tangentially driven polymers in free space (dotted lines). $\tau_{e,0}$ is the time at which the autocorrelation of the unit end-to-end vector $\hat{\mathbf{r}}_e$ decays to $1/e$. Standard errors are within the marker size in (a), and the standard deviation is shown as the shaded region in (b).}
\label{fig:Simulations}
\end{figure}

In this study, we use \textit{T.~Tubifex} worms, which behave like active polymers \cite{Deblais2020a,Deblais2020b,Heeremans2022,Deblais_worm_review_2023}, to investigate locomotion in quasi-2D porous media with cylindrical pillars. We compare their dynamics with a computational model of tangentially driven active filaments, commonly used for self-propelled biopolymers \cite{Mokhtari2019,winkler_physics_2020,Fazelzadeh2023}. Worm transport depends on the surface fraction $\phi$ occupied by pillars and obstacle order. In disordered cylindrical pillar arrays, $D_l$ increases with $\phi$, whereas in square lattices, it decreases. This transport enhancement arises from worms navigating curvilinear tubes formed by randomly positioned pillars, contrasting with their entrapment in ordered configurations. Surprisingly, lower activity enhances spreading in disordered media by increasing effective persistence length, promoting longer trajectories. Simulations confirm that tangentially driven active filaments~\cite{IseleHolder2015} reproduce the long-time dynamic trends of \textit{T.~Tubifex} worms.

When deposited on a free surface, the wiggling motion of the \textit{T.~Tubifex} enables the worm to crawl on the surface, giving rise to a diffusive motion. The crawling motion is effectively two-dimensional, since the worms are denser than water and, therefore, always confined to the bottom of the geometry (Sup.~Vid.~1).

In our experiments, a single \textit{T.~Tubifex} worm is introduced into cylindrical pillar arrays with radius $R_p = 2.5$ mm on a 2D square surface (44$\times$44 cm$^2$), immersed in thermostated water at $T$ = 21$^{\degree}$C. Worm motion is tracked for 2 hours by taking a sequence of pictures. We investigate two distinct pillar arrangements: a periodic square lattice and a disordered array, where $N$ static cylindrical pillars were randomly distributed with a minimum distance of 0.5 mm between them to allow worm passage (see Sec.~I and Figs.~S1 and S2 in the Supplementary Materials~\cite{Sup}). The pillar surface fraction was varied from $0.1$ to $0.6$. All reported conformational and dynamical features of the worms at $T$ = 21$^{\degree}$C were averaged from trajectories of at least 30 worms with a contour length $L_c = 23 \pm 8$ mm, ensuring the length range had no significant impact on the results (Sup.~Fig.~S3, \cite{Sup}).

\begin{figure}
\centering
\includegraphics[width=1\columnwidth]{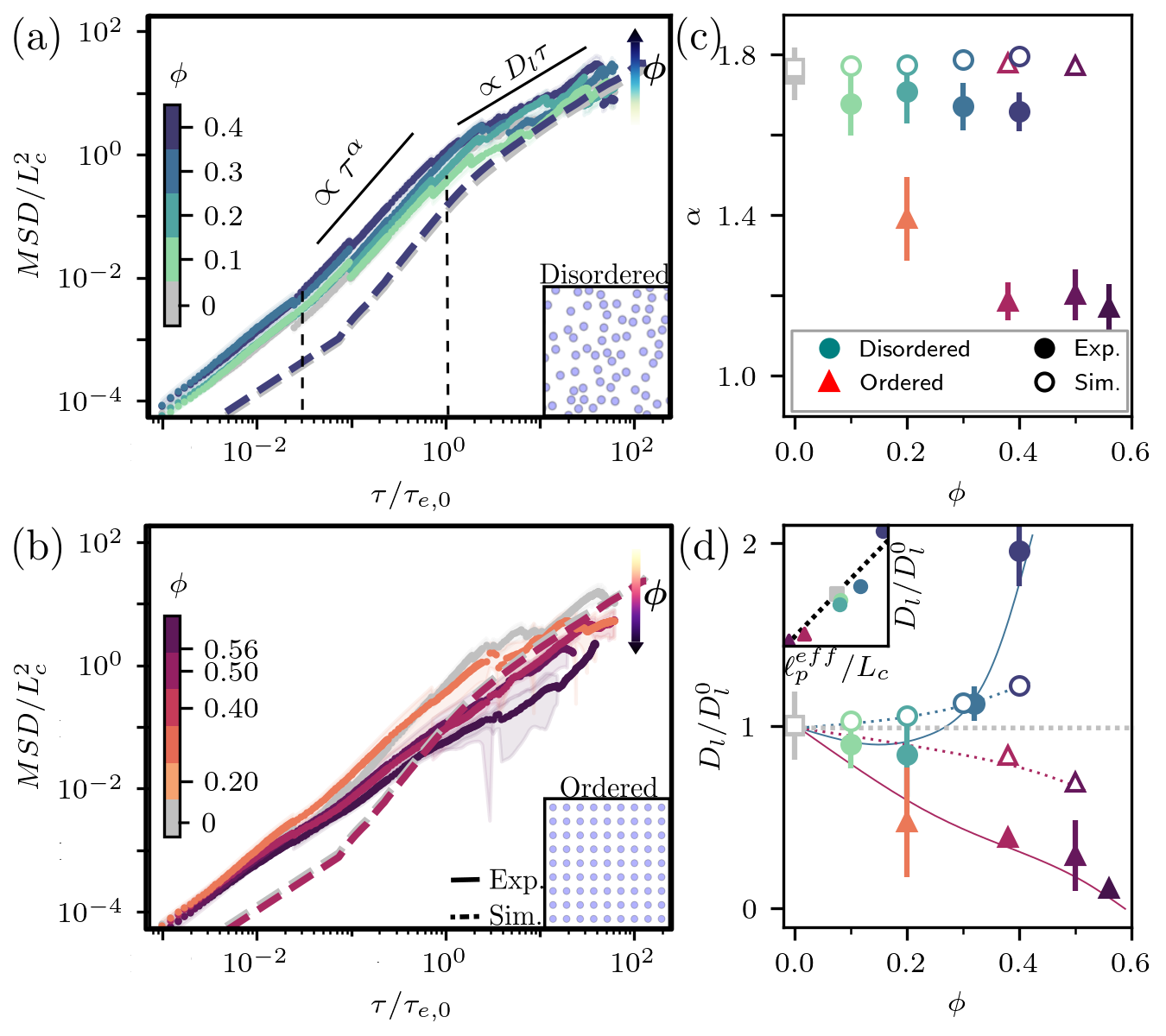}
\caption{Mean square displacements (MSD) normalized by contour length $l_c$ and time normalized by reorientational time $\tau_{e,0}$ for living worms in disordered (a) and ordered (b) obstacle arrangements at $T$ = 21$^{\degree}$C. In disordered media (a), MSD increases faster with $\phi$, while the opposite is observed in ordered media (b). Dashed lines show simulations of the tangentially driven model for $\phi$ = 0 (free) and 0.4. (c) Slopes $\alpha$ of MSD curves for living worms (filled) and active polymer simulations (open). (d) Long-time diffusion coefficient $D_l$ normalized by $D_l^0$ as a function of $\phi$, with lines guiding the eye. The inset shows the linear relationship between $D_l$ and $\ell_p^{eff}/L_c$. Standard errors are shown as shaded regions in (a),(b) and error bars in (c),(d).}
\label{fig:MSD}
\end{figure}

\begin{figure*}
    \centering
    \includegraphics[width=0.8\textwidth]{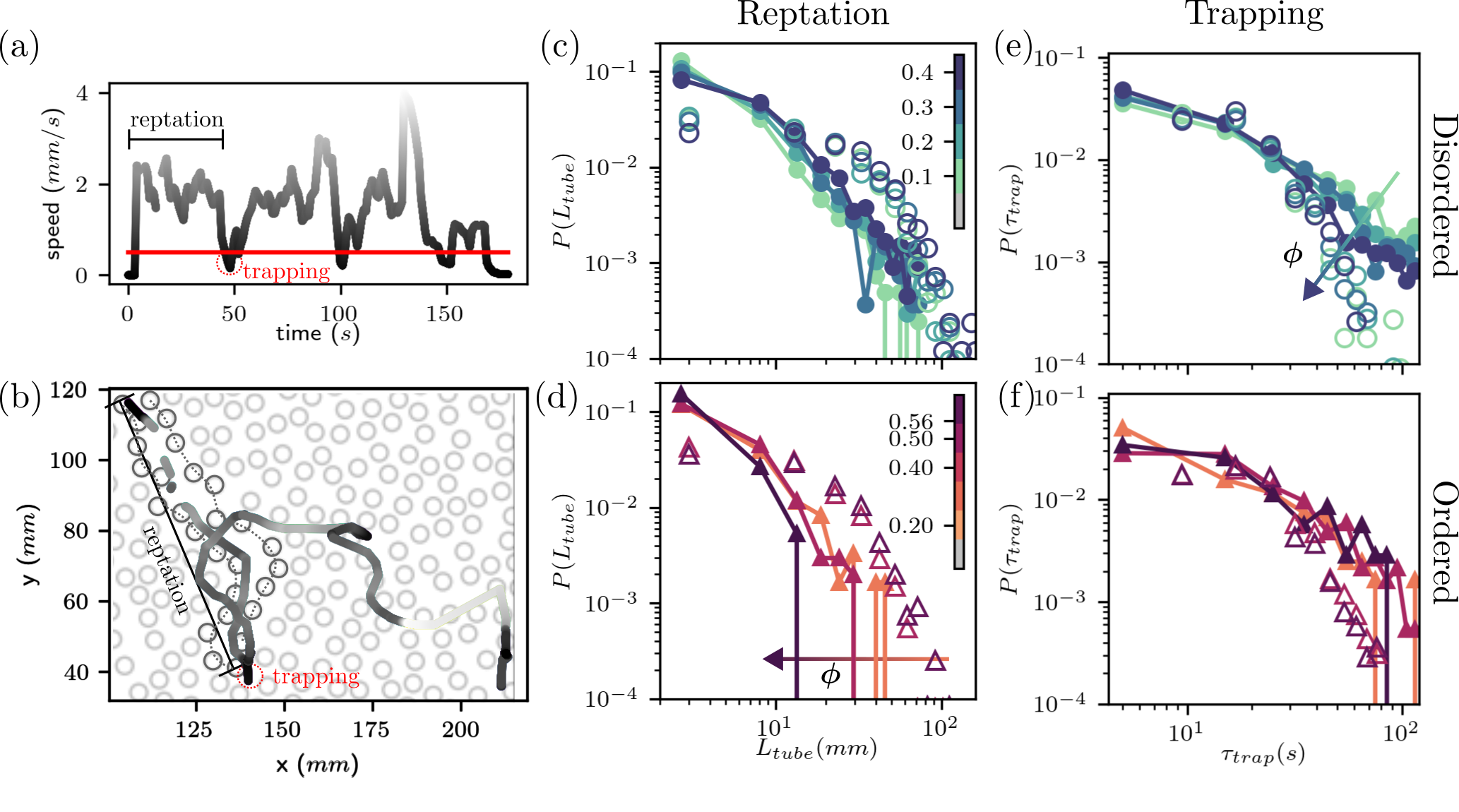}
\caption{(a) Example of a worm trajectory ($\phi=40\%$) as it reptates within a tube formed by disordered pillar positions. (b) Worms are caged when their instantaneous speed drops below $0.5~mm/s$. Between caging events, they crawl through effective tubes of length $L_{tube}$. (c),(d) Tube length distributions for disordered (c) and ordered (d) media. In the disordered case, longer tube lengths are observed with increasing $\phi$, whereas in the ordered case, maximum tube length decreases at higher $\phi$. The opposite trend is seen for trapping time distributions in disordered (e) and ordered media (f). Open symbols represent simulation results.}
    \label{fig:runlength}
\end{figure*}

In Figure~\ref{fig:Trajectories}, we show typical trajectories of the center-of-mass of a worm moving in a disordered~(a) and ordered~(b) pillar arrays (see also Sup.~Vid.~2 \& 3). In both geometries, as time progresses, the worm is able to navigate through the obstacles but their dynamics are found to be dependent on the pillars arrangement and density.
To investigate the similarities with active polymers, we conducted simulations using the tangentially driven polymer model, as described in~\cite{IseleHolder2015,Fazelzadeh2023} (see also Sec.~II in Sup.~Mat.~\cite{Sup}). The motion of each monomer follows overdamped Langevin dynamics, including the active force of amplitude $f^a$ per monomer along the tangent of the backbone of the polymer and the bending stiffness $\kappa$ between neighboring bonds, which represents the inherent flexibility of the active polymer. The key parameters of the model thus boil down to ($f^a$, $\kappa$).
To compare the tangentially driven polymers to the worms we measure their tangent-vector orientational correlations and average over time (Sup.~Fig.~S4\&~S5, \cite{Sup}). From this we extract the effective persistence length $\ell_p^{eff}$, which correlates to $\kappa$.
The effective persistence length $\ell_p^{eff}$ is shown by bold symbols in Fig.~\ref{fig:Simulations}(a) as a function of $\phi$ in ordered and disordered media, which shows an increase (decrease) with $\phi$ for worms in ordered and disordered media.
Then, we adjust the parameter $\kappa$ in our simulations to match the observed $\ell_p^{eff}$, see the open symbols in Fig.~\ref{fig:Simulations}(a).

The active force $f^a$ is selected to ensure it dominates at the level of the whole polymer while still allowing biological fluctuations to affect the monomer's dynamics (Sup.~Vid.~4). To compare experiments and simulations, we rescale the time by the reorientational relaxation time of the worm in free space, determined from the $(1/e)$-decay time of the autocorrelation function of the end-to-end vector, see Fig.~\ref{fig:Simulations}(b). For tangentially driven polymers in free space with sufficiently large active force ($f^a > 0.01$), both longtime diffusion $D_l$, dominated by activity, and the orientational relaxation time of the end-to-end vector ($\tau_{e,0}$) scale linearly with active force per monomer ($f^a$). Therefore, rescaling the time by $\tau_{e,0}$ eliminates the explicit dependence on the active force, making the precise value of $f^a$ less critical~\cite{Fazelzadeh_23_PRE,Bianco2018}.
 
We examine the effect of the geometry of the obstacle arrays and the concentration of the pillars on the motion of a self-locomoting worm by measuring the mean squared displacement (MSD). Figure~\ref{fig:MSD}(a) and (b) show MSDs as functions of rescaled time $\tau/\tau_{e,0}$, where the worm's reorientational relaxation time in a free environment $\tau_{e,0} \approx$ 40~s. 

For short time scales ($\tau/\tau_{e,0} \leq 0.02$), the worms have not yet encountered the pillars, resulting in only slight superdiffusive behavior with a diffusion exponent of 1.2. After this typical time scale, our living worms interact with the pillars with a dynamics that depends on the degree of environmental order. In the disordered case, we observe an almost ballistic regime as shown by $\alpha \sim 1.8$ in Fig.~\ref{fig:MSD}(a), which we found to be almost independent of the surface fraction of the pillar $\phi$ [Fig.~\ref{fig:MSD}(c)]. At long times $\tau/\tau_{e,0} \gg 1$, each MSD curve exhibits a diffusive regime where $MSD=4 D_l t$, with a slope that surprisingly increases when $\phi \geq 0.2$. For the lowest $\phi$, the worms do not experience significant confinement, leading to minimal differences in diffusion compared to the obstacle-free case.
For the ordered media [Fig.~\ref{fig:MSD}(b)], the opposite trend is observed: We observe a weak superdiffusive regime and almost a direct crossover from short-time dynamics to the diffusive regime with a slope that decreases with increasing $\phi$. 

As a result, the presence of an intermediate regime of super-diffusion depends on the arrangement of obstacles, whether they are ordered or not.

The spatial arrangement and density of obstacles influence the long-time diffusion coefficient $D_l$ of the active worms. As shown in Fig.~\ref{fig:MSD}(d), the two geometries exhibit opposite trends: In disordered media, $D_l$ increases with $\phi$, while it decreases in ordered media, obeying the intuitive expectation that greater crowding slows down the diffusion. Interestingly, the time scales $\tau \sim \tau_{e,0}$, at which we observe the diffusive regime, do not appear to depend on the obstacle surface fraction $\phi$. Moreover, extracting $\tau_e(\phi)$ from the orientational-time autocorrelation of the end-to-end vector (see Sup.~Fig.~S6 in \cite{Sup}) confirms that $\tau_e$ is independent of $\phi$, consistent with simulation results for semiflexible active polymers in periodic obstacle arrays~\cite{Fazelzadeh2023}.

Examining the worm's center-of-mass trajectories in ordered and disordered pillar arrays (Figs.~\ref{fig:Trajectories}(a),(b)) reveals that distinct locomotion strategies drive the contrasting diffusion trends. In disordered media, worms navigate curvilinear tubes formed by randomly positioned pillars, intermittently trapping before switching paths, akin to reptation in crowded environments~\cite{deGennes1971,doi1986theory,Muthukumar1989}. During trapping, worms coil up in pores (Fig.~\ref{fig:Trajectories}(b)). In contrast, in ordered structures, worms appear more flexible, lingering longer in pores and occasionally hopping between them, consistent with the lower persistence length in Fig.~\ref{fig:MSD}(e).  

The distributions of curvilinear tube lengths and trapping times reflect these behaviors. Trapping events occur when the center-of-mass speed drops below $v = 0.5$~mm/s, localizing the trajectory [Figs.~\ref{fig:runlength}(a),(b)]. The tube length $L_{tube}$ is the distance traveled between traps. Probability distributions $P(L_{tube})$ [Figs.~\ref{fig:runlength}(c),(d)] show that in disordered media, higher pillar densities lead to longer runs, increasing $P(L_{tube})$ in the tail [Fig.~\ref{fig:runlength}(c)]. In contrast, in ordered media, increasing $\phi$ shortens runs, reducing the tail of $P(L_{tube})$ [Fig.~\ref{fig:runlength}(d)].  

Next, we analyze the trapping time distribution $\tau_{trap}$, defined as the duration of trapping events. Figs.~\ref{fig:runlength}(e),(f) show $P(\tau_{trap})$ for disordered and ordered media. In disordered media, the tail of $P(\tau_{trap})$ decreases with $\phi$, indicating longer tube residence times [Fig.~\ref{fig:runlength}(e)], while in ordered media, it remains nearly unchanged [Fig.~\ref{fig:runlength}(f)]. Our simulations of tangentially driven filaments capture the general shape of $P(\tau_{trap})$ but not its exact dependence on $\phi$, warranting further investigation. The ordered geometry at $\phi = 0.5$ forms a slightly defective hexagonal lattice, resulting in a non-monodisperse void size distribution with a minor log-normal contribution (Sup.~Fig.~S7).  

The contrasting trends in the long-time diffusion coefficient and the distribution of run lengths within tubes on the pillar packing fraction $\phi$ for ordered and disordered media originate from changes in the effective persistence length of the worms with increasing $\phi$. The active contribution to long-time diffusion ($D_l^a$) is dominant, and theoretical calculations for tangentially driven active chains predict that it scales as 
\begin{equation}D_l^a = {f^a}^2 \langle R_e^2 \rangle \tau_e / (\gamma^2 L_c^2), \label{eq:DL}
\end{equation}
irrespective of $\phi$ \cite{Fazelzadeh2023}, also valid in free space \cite{Fazelzadeh_23_PRE,Bianco2018} (with the average end-to-end distance $\langle R_e^2 \rangle$ and friction coefficient $\gamma$). Additionally, for sufficiently flexible active polymers with $ \ell_p/L_c \ll 1$, the relation  $\tau_e \sim L_c / f^a$  holds~\cite{Fazelzadeh_23_PRE,Bianco2018} leading to $D_l^a \sim f^a \langle R_e^2 \rangle / L_c$. For the worms $ 0.1<\ell_p/L_c <0.26 $ is sufficiently small to satisfy this relationship, while the contour length $L_c$ and the friction coefficient $\gamma$ do not change. Experimentally, measured values of $\tau_e$ exhibit a weak dependence on $\phi$, see Fig.~S6 (b) and (c) but their value is systematically higher for disordered arrays in comparison to ordered ones.

Assuming $\tau_e$ is roughly independent of $\phi$ but higher in disordered than ordered obstacle arrangements, the product $\tau_e \langle R_e^2 \rangle$ explains the observed long-time diffusion trends. 
Estimating $\langle R_e^2 \rangle$ experimentally is challenging, but using the worm-like chain model for semiflexible worms, we get $\langle R_e^2 \rangle = 2 \ell_p^{\text{eff}} L_c$ for $\ell_p^{\text{eff}}/L_c \ll 1$. Thus, we expect $D_l \sim \ell_p^{\text{eff}}$, as confirmed in Fig.~\ref{fig:MSD}(d) (inset). In disordered media with $\phi \geq 0.3$, increased reorientation time and extended conformations enhance diffusion, surpassing that of free particles.

Although the tangentially driven polymer model captures long-time dynamics, it does not fully describe aspects like the anomalous diffusion exponent in the intermediate MSD regime [Fig.~\ref{fig:MSD}(c)]. Achieving full agreement with experiments requires models incorporating heterogeneous or time-dependent active forces or transversal motion modes \cite{Vatin2024,anand_structure_2018,anand_conformation_2020,prathyusha2022emergent}. A detailed worm–polymer model comparison, using principal component analysis of contour curvature \cite{Saggiorato2017}, is provided in Sup.~Sec.~IID and Sup.~Fig.~S14.

\begin{figure}[t]
    \centering
    \includegraphics[width=\columnwidth]{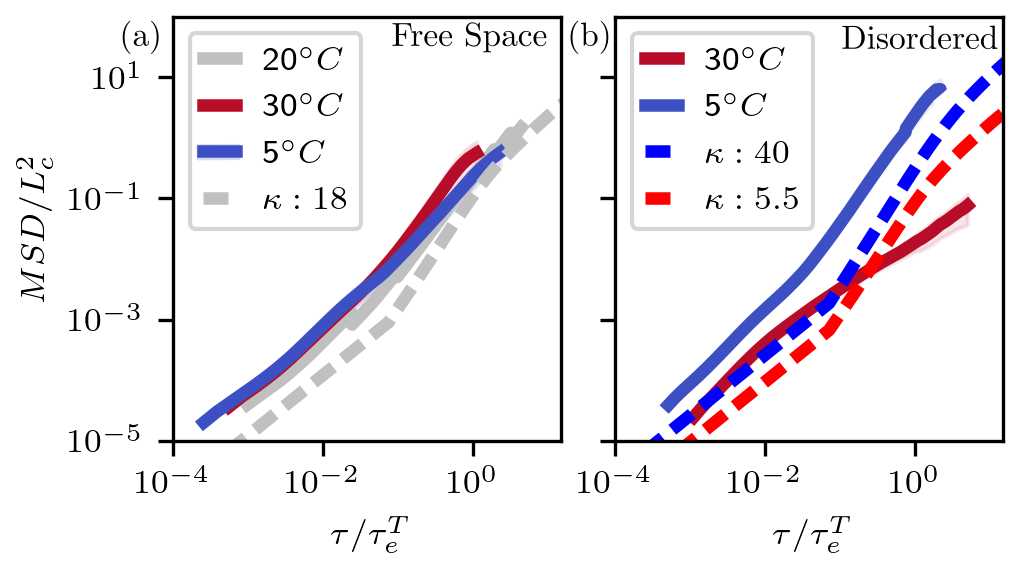}
  \caption{(a) MSDs of \textit{T.~Tubifex} worms at high ($T = 30^{\circ}$C, red line), low ($T = 5^{\circ}$C, blue line) and intermediate ($T = 21^{\circ}$C, silver line) activity levels. Time is normalized by the respective reorientational timescales ($\tau_e^{5\celsius} = 78$ s, $\tau_e^{21\celsius} = 41$ s, $\tau_e^{30\celsius} = 36$ s), which are temperature dependent (Sup.~Fig.~S9~\cite{Sup}). Dashed lines show tangentially driven polymer simulations with matching persistence length and $f^a = 0.1$. (b) MSDs of worms at high and low temperatures in disordered porous media. At lower temperatures, the effective persistence length increases, enhancing worm spread compared to high temperatures ($T = 30^{\circ}$C, red line). Dashed lines correspond to simulation results adjusted for $\kappa$.}
\label{fig:EffectofActivity}
\end{figure}

Finally, we examine the impact of activity on worm transport, adjustable via ambient water temperature \cite{Deblais2020b}. The average persistence length of worms remains unchanged with temperature \cite{Heeremans2022}, whereas their long-time diffusion coefficient ($D_l$) increases, and their reorientational time ($\tau_e$) decreases with temperature \cite{Deblais2020b}. This behavior aligns with tangentially-driven active polymers, where $D_l$ scales linearly with active force ($f^a$) and $\tau_e$ inversely scales with it, $\tau_e \sim 1/f^a$ \cite{IseleHolder2015,Bianco2018,Fazelzadeh_23_PRE}. Thus, we expect the mean squared displacement (MSD) versus time, scaled by $\tau_e^T$, to be temperature-independent. Fig.~\ref{fig:EffectofActivity}(a) shows the MSD for worms in free space at three activity levels (low, T = $5\celsius$, intermediate, T= $21\celsius$, and high, T = $30\celsius$). While higher temperatures result in larger MSDs, see Sup.~Fig.~S8, \cite{Sup} for the same data in lab-units), normalizing by $\tau_e^T$ largely mitigates this effect leading to collapse of data of different temperatures.

In disordered environments, however, the MSD versus rescaled time ($\tau/\tau_e^T$) strongly depends on activity level, as shown in Fig.~\ref{fig:EffectofActivity}(b). Surprisingly, worms with lower activity (T = $5^\circ$C) exhibit faster long-time diffusion than those with higher activity (T = $30^\circ$C), despite the increase in $\tau_e$ at lower temperatures. This suggests that the contrasting trend is not due to increased worm activity in the porous medium at lower temperatures. Interestingly, the effective persistence length of worms in disordered media, averaged over the experimental timescale (Sup.~Fig.~S9 \cite{Sup}), increases from $\ell_p^{eff}$ = 0.12 to 0.6 by lowering the temperature from T = $30^\circ$C to $5^\circ$C.

To rationalize these experimental findings, simulations are conducted in the same geometry (Sup.~Vid.~5 \& 6). The active force is kept constant, and bending stiffness values are chosen to match the effective persistence length of worms ($T=30\celsius: \kappa = 5.5$, $T=5\celsius: \kappa = 40$, dashed lines). At low activity, simulations of tangentially driven polymers align well with worm behavior at large timescales. However, at higher temperatures, the model no longer fully captures worm dynamics, suggesting that under low temperatures and confinement by pillars, worms exhibit peristaltic-like motion. At higher activity, worms adopt transversal motion modes, indicating the need for a more refined model.

In conclusion, our study highlights the crucial role of environmental heterogeneity, porosity, and the interplay between flexibility and activity in shaping active flexible agents’ dynamics, resulting in non-trivial effects distinct from passive systems \cite{Hofling2008,hofling2013}. Unlike previous studies on stiff and semi-flexible polymers \cite{Mokhtari2019,Bhattacharjee2019,theeyancheri_active_2023}, we show that pore morphology significantly impacts worm spreading. For $\ell_p/L_c \sim 0.1$, disordered cylindrical pillars enhance long-time diffusion with increasing packing fraction, while ordered arrays reduce it. Worm persistence length depends on obstacle arrangement and packing fraction. Theory~\cite{Fazelzadeh2023} and simulations indicate that $D_l$ is governed by how porous media alter worms’ mean conformation (e.g., curling) and reorientation time. In disordered media, enhanced diffusion arises from curvilinear tubes increasing both mean reorientation time and end-to-end distance. Long-time diffusion follows Eq.~\eqref{eq:DL}, linking it to the product of mean squared end-to-end distance and reorientation time.

While perhaps not universal, our findings emphasize cylindrical pillar order, with curvilinear tube formation and increased persistence length influence spreading. Similar effects appear in ordered square lattices (Sup.~Fig.~S15). While these results hold, further studies on diverse geometries (e.g., hexagonal lattices) are needed for broader generalization. These results suggest new ways to control active polymers via lattice arrangement and pillar shape. Additionally, our observations indicate that distributing active forces along a driven polymer could enhance active transport, as seen in living worms. Future work could explore worm locomotion in more complex, heterogeneous environments with stagnant regions, offering insights into their behavior in realistic settings.

\bibliography{refs}
\end{document}


\title{Supplementary Materials for\\ ``Locomotion of Active Polymerlike Worms in Porous Media''}

\author{R.~Sinaasappel}
\affiliation{Van der Waals-Zeeman Institute, Institute of Physics, University of Amsterdam, 1098XH Amsterdam, The Netherlands.}
\author{M.~Fazelzadeh}
\affiliation{Institute for Theoretical Physics, University of Amsterdam, Science Park 904, 1098XH Amsterdam, The Netherlands.}
\author{T.~Hooijschuur}
\affiliation{Van der Waals-Zeeman Institute, Institute of Physics, University of Amsterdam, 1098XH Amsterdam, The Netherlands.}
\author{Q.~Di}
\affiliation{Institute for Theoretical Physics, University of Amsterdam, Science Park 904, 1098XH Amsterdam, The Netherlands.}
\author{S.~Jabbari-Farouji}
\email{s.jabbarifarouji@uva.nl}
\affiliation{Institute for Theoretical Physics, University of Amsterdam, Science Park 904, 1098XH Amsterdam, The Netherlands.}
\author{A.~Deblais}
\email{a.deblais@uva.nl}
\affiliation{Van der Waals-Zeeman Institute, Institute of Physics, University of Amsterdam, 1098XH Amsterdam, The Netherlands.}

\date{\today}

\begin{abstract}
This Supplementary Material provides additional information on the experimental setup and methodologies employed in this study and on the tangentially driven polymer model utilized for our analysis.
\end{abstract}

\pacs{Valid PACS appear here}
\maketitle

\section{EXPERIMENTS}
\subsection{Experimental set-up}

In our experiments, we placed a single \textit{T.~Tubifex} worm into pillar arrays submerged in a thermostated water volume, tracking their motion in real-time through 2-hour video recordings. We investigated two geometries: (i) a periodic crystalline structure and (ii) a disordered geometry with randomly positioned pillars.

\begin{figure}[]
    \centering
    \includegraphics[width=0.9\textwidth]{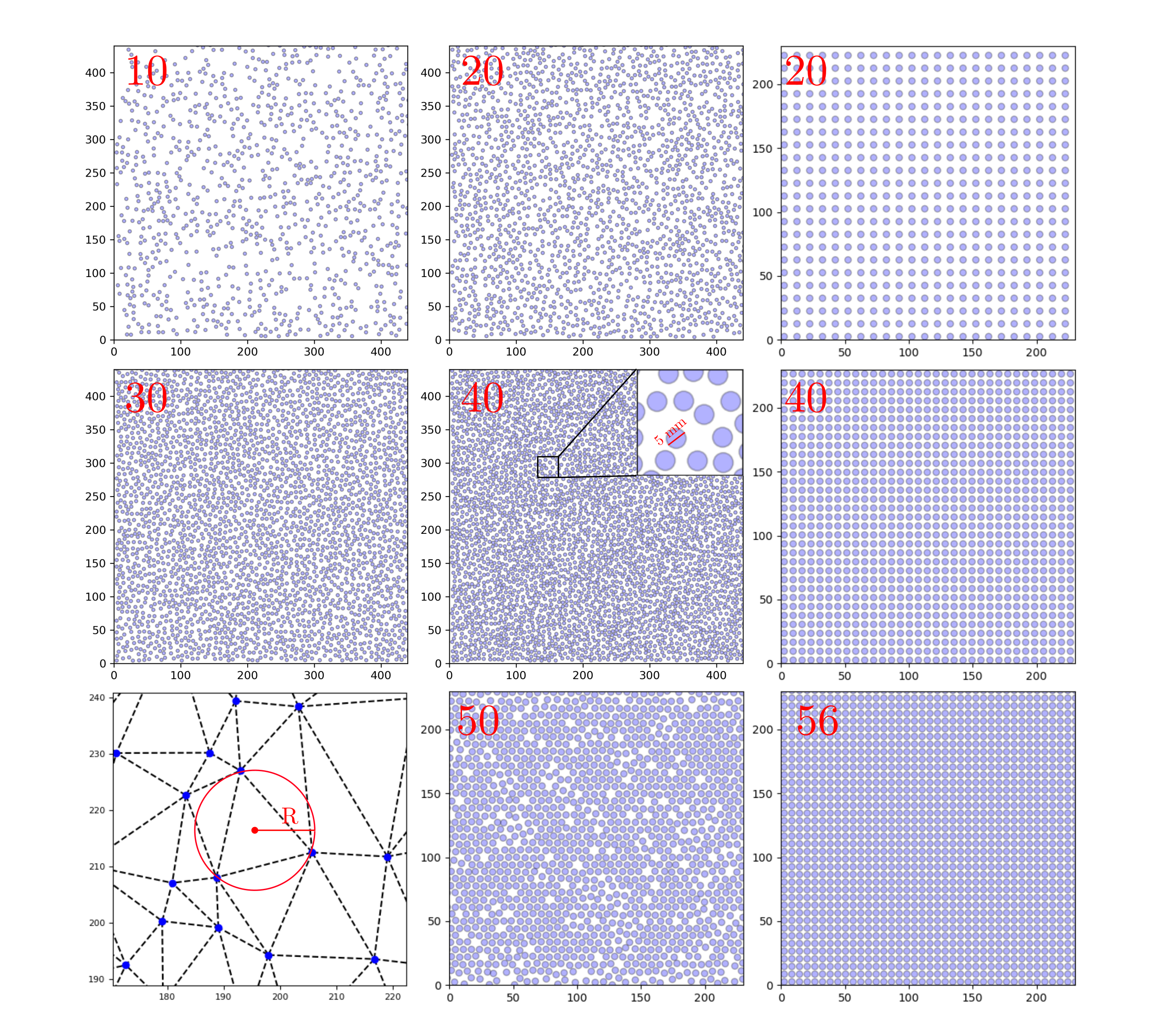}
    \caption{\textbf{The ordered and disordered geometries used in the experiments.} The axis are in millimeters and the radius of the pillars is $r$ = 2.5 mm. The surface fraction occupied by the pillars is indicated in the top left corner in red. The bottom left corner shows the Delaunay triangulation.}
    \label{SupFig:mazes}
\end{figure}

For each pattern, we inserted $N$ static pillars with a radius $R_{p}$ = 2.5 mm on a square two-dimensional surface of the same material, ensuring no overlap between the pillars. We imposed the condition that the minimum distance between two pillars is approximately the characteristic width of a worm ($\sim$ 500 $\mu$m) to allow worm passage. The surface fraction of the pillar varied from $10\%$ to $60\%$, calculated as $\phi = N \pi {R_p}^2 / L^2$. In the disordered medium, pillar positions were randomly selected using the \textit{numpy.random} library in Python. Pillar placement continued sequentially, retrying if a selected location was closer than 2.5+1 mm to a previously placed pillar, until reaching the desired surface fraction $\phi$. Supplementary Figure~S\ref{SupFig:mazes} shows the experimental geometries, with dimensions $L^2$ of 230x230 mm$^2$ for ordered and 440x440 mm$^2$ for disordered setups. The larger setup involved cutting holes in an acrylic sheet and placing short acrylic rods with a radius of 2.5 mm; the ordered geometries were 3D-printed. We observed no significant differences in the behavior of the worm between the two dimensions. 

\begin{figure}[]
    \centering
    \includegraphics[width=0.9\textwidth]{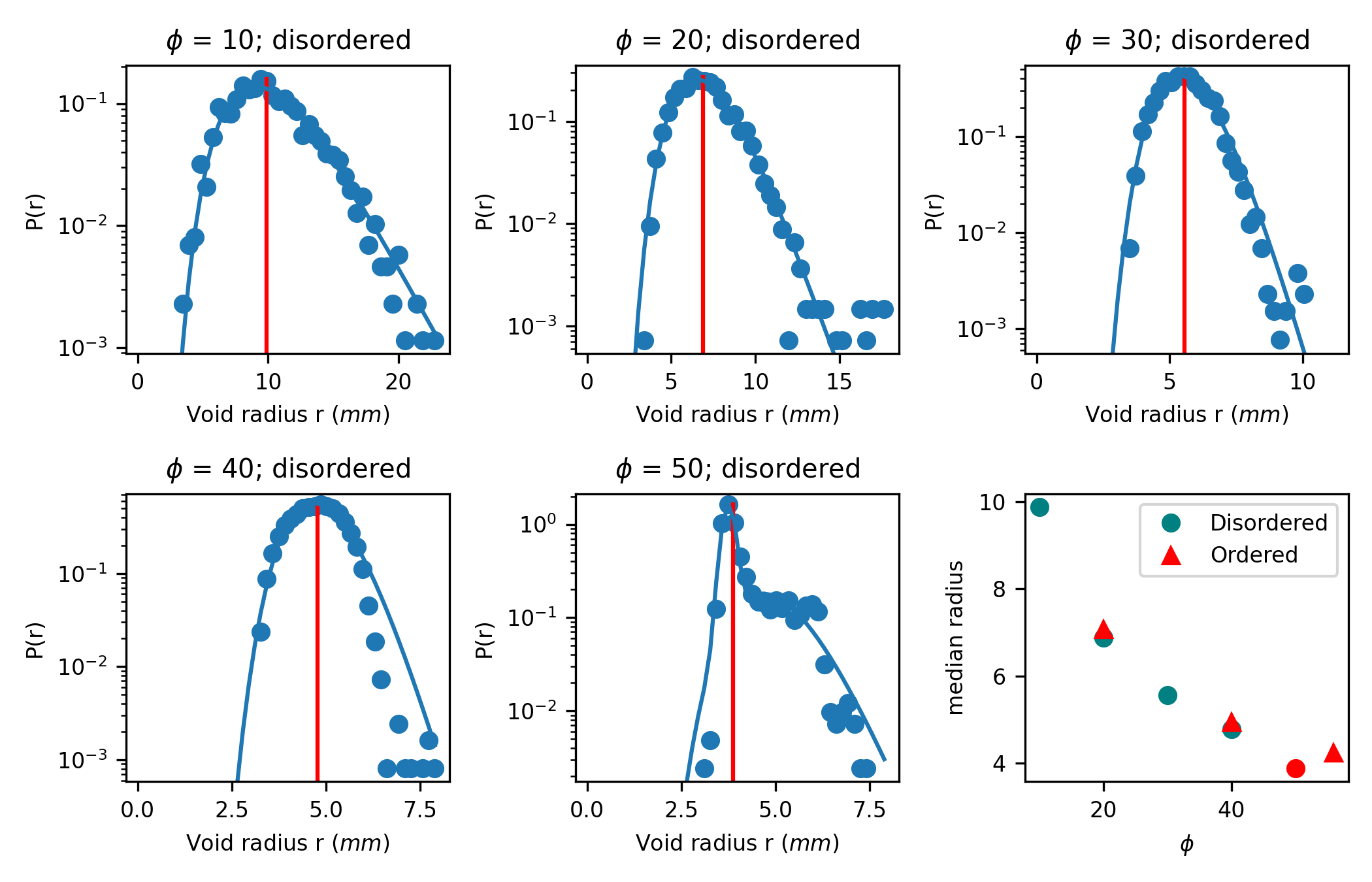}
    \caption{\textbf{Pore size distribution in the geometries.} Distribution of the radii of all possible biggest circles that do not contain any points, as found through Delaunay triangulation, as a measure for the size of the voids in the pillar arrays. The distribution of the void sizes are well fitted with a log-normal distribution. For $\phi$ = 50\% the distribution is fitted by the sum of a Gaussian and a log-normal distribution.}
    \label{SupFig:cavity_dist}
\end{figure}

To quantify the size of the voids in the different geometries, we used Delaunay triangulation (as implemented in the \textit{scipy.spatial} library). Delaunay triangulation divides an area containing a set of points into triangles, such that the circumcircles of these triangles do not contain any points. It allows us to find the biggest possible circles that one can draw that do not contain any of the pillars. The radius of these circles are taken as a good approximation for the void sizes in the geometries, so it is possible to compare how the void size distribution changes with the surface fraction of the pillars. Supplementary Figure~S\ref{SupFig:cavity_dist} shows the distribution of void radii in the pillar arrays, following a log-normal distribution. For larger surface fraction above $\phi=50\%$, we were unable to produce a disordered medium using the standard method. Instead, we started from an evenly spaced hexagonal lattice, randomly removing pillars to achieve the desired surface fraction. Next, the pillars where allowed to diffuse for a while. This resulted in a (largely) hexagonal lattice with defects, exhibiting a more mono-disperse void size distribution. Due to this, the worms behaved similarly to those in the square lattice geometries, therefore they will be referred to as ordered in subsequent materials.


\subsubsection{Tracking and mean square displacement (MSD)}

In the experiments, worms are positioned atop the geometries in a 15 cm deep water bath, dimly illuminated from below with an LED panel. The camera recordings from above capture the experiments (Nikon D5300 equipped with a macrolens), which are subsequently analyzed using a Python script. From the images, we extracted the center of mass (CoM) and the contour of the worm, as shown in Figure~S\ref{fig:Exp_PersistenceLength}. 

After tracking the worm's CoM position ($\textbf{r}_{cm}=(x,y)$), we compute the mean square displacement as a function of lag time $\mathrm{MSD}(\tau) = \langle (r_{cm}(\tau)-r_{cm}(0))^2 \rangle$. Each experiment yields one MSD curve, which is then averaged to produce the curves reported in the main paper. Supplementary Figure~S\ref{fig:all_msd} displays the MSD curves of all experiments with the average represented by the black line.
\clearpage
\newpage
\subsection{Effect of worm's contour length, $L_c$}
\begin{figure}[h]
    \centering
    \includegraphics[width=\textwidth]{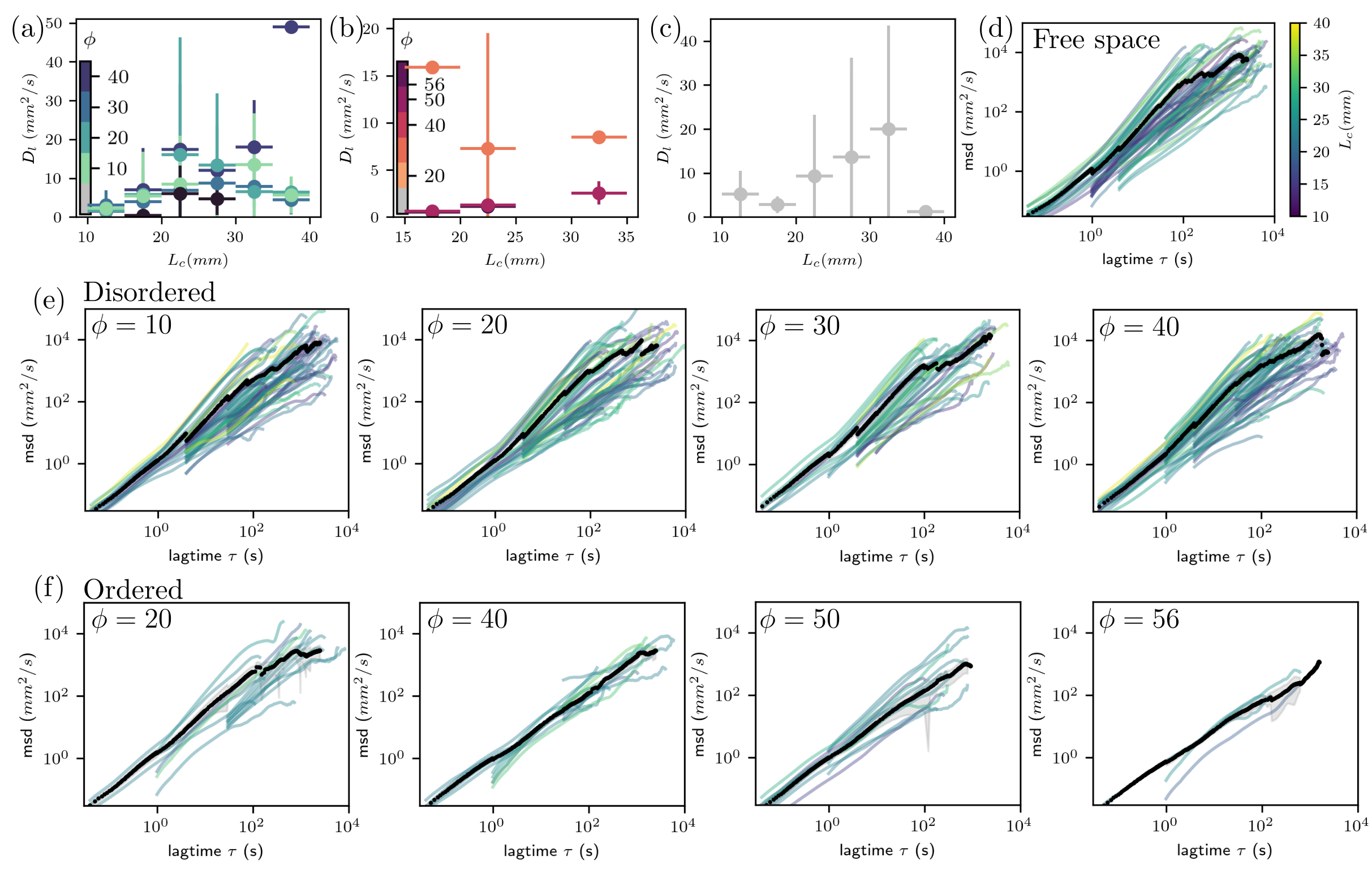}
    \caption{\textbf{MSD curves of all experiments.} (a) Long-time diffusion constant versus the contour length of worms in the disordered, (b) ordered media, and (c) in free space. (d) MSD curves of worms in free space in lab units. The color of the lines indicates the contour length of the worms. The average is indicated by the black line. (e) All MSD curves for the disordered medium. The color of the lines is mapped according to the color bar in (d). The surface fraction of the pillars is indicated in the top left corner. (f) All MSD curves for the ordered medium. The color of the lines is mapped according to the color bar in (d). The surface fraction of the pillars is indicated in the top left corner.}
    \label{fig:all_msd}
\end{figure}

\begin{figure}[h]
\centering
\includegraphics[width=0.6\textwidth]{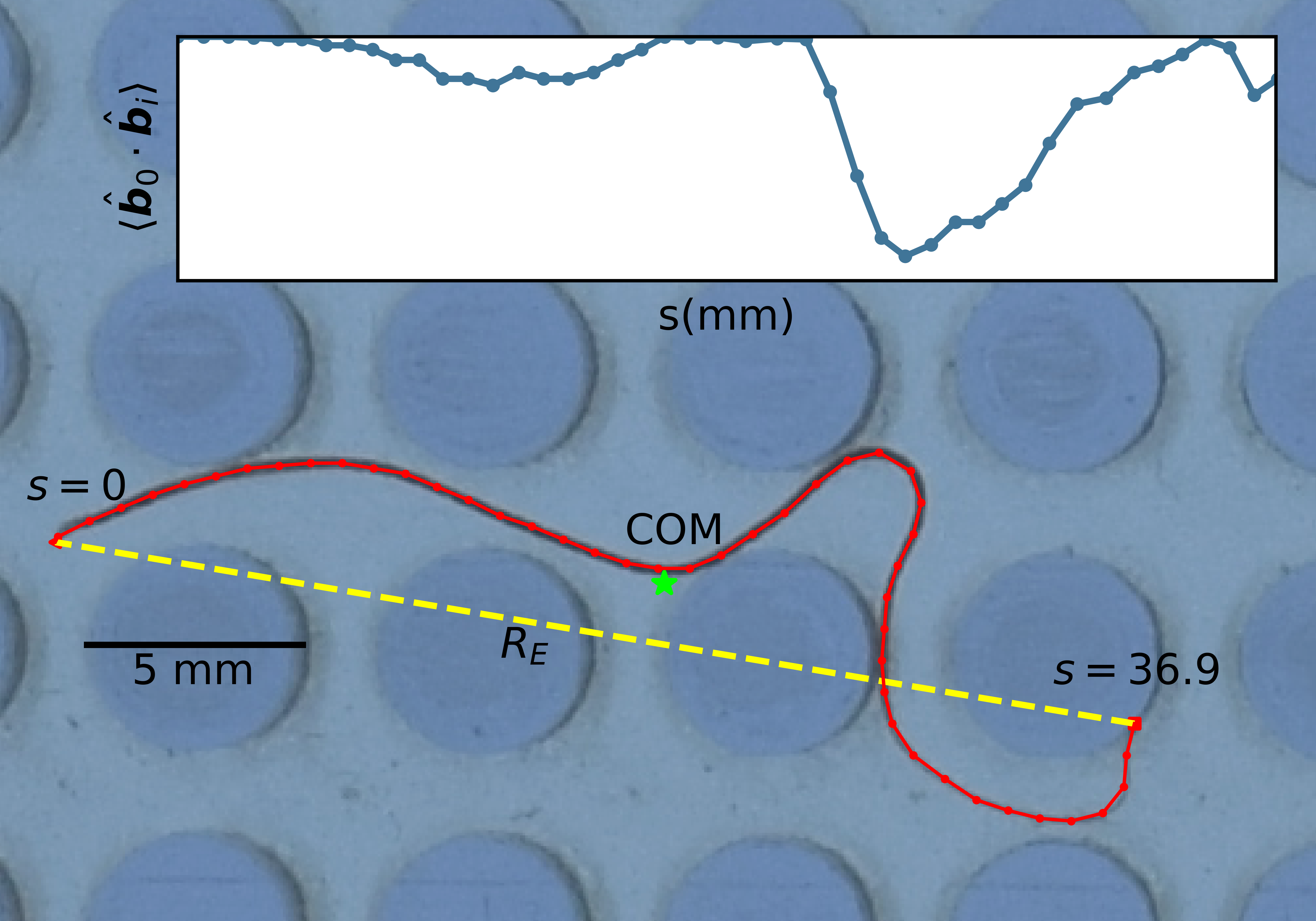}
\caption{\textbf{Determination of the persistence length}. } 
\label{fig:Exp_PersistenceLength}
\end{figure}

In our experiments, worms of varying lengths were used due to experimental constraints (batch of worms are polydisperse in their contour lengths). However, we did not observe any significant correlation between the length of the worms and their behavior. Analysis of the long-time diffusion constant and MSD curves revealed no discernible correlation with the length of the worm, as illustrated in Figure~\ref{fig:all_msd}(a-f), where the color code represents the contour length of the worm.

\clearpage
\newpage
\subsection{Conformation}

\subsubsection{Effect of the geometry and temperature on the persistence length}
To determine the appropriate bending stiffness $\kappa$ of the semi-flexible polymers in the simulations, we calculated the effective persistence length $\ell_p^{eff}$ from the videos of the worms for different surface fractions and in ordered and disordered media. Calculating the persistence length involves defining an effective bond vector for the worms. Firstly, we remove the background from each image and select a smaller region of interest (ROI) of the total image where the worm is either 600x600 or 360x360 pix$^2$, depending on the total image size. Within this selected image region, we identify all pixels belonging to the individual worm and employ a skeletonization algorithm to obtain a single-pixel-wide chain of pixels, which forms the initial polymer-like backbone of the worm. Each pixel can be regarded as a monomer, with neighboring pixels connected by bond vectors $\textbf{r}_{i,j}$.

However, at the single-pixel level, each neighbor has only 8 possible directions it can be connected, resulting in discrete bond vectors and sharp lines in the bond-bond correlation function. Additionally, at high resolution, each pixel represents a minute scale, rendering the bonds effectively rigid, leading to a plateau in the bond-bond correlation. To address these issues, we average every four pixels to create a new monomer \( \textbf{r}_i^* \), and these are connected by new bond vectors, denoted \( \textbf{b}_i \), where \( i \) ranges from 0, the first bond in the chain, to \( i = N-1 \), the last bond.

\begin{figure}[H]
\centering
\includegraphics[width=0.9\textwidth]{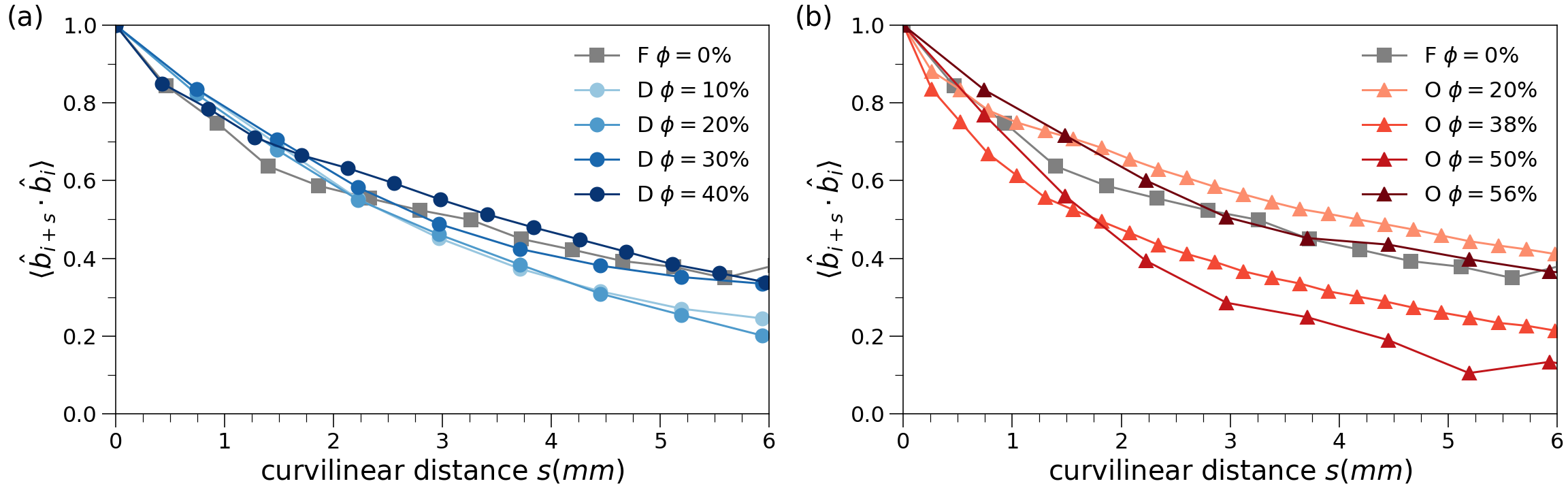}
\caption{\textbf{Effect of the lattice pattern on the persistence length.} (a) Bond-Bond correlation of the worms in the disordered medium. The persistence length increases with increasing the packing fraction (b) The effect of the increasing packing fraction in a ordered arrangement on the persistence length. The bond-bond correlation curves are averaged in time over around 10 minutes of footage per worm for 5 different worms.}
\label{fig:PersistenceLength}
\end{figure}

To determine the bond-bond correlation function, we examine the dot product between a normalized bond vector \( \hat{\textbf{b}}_i \) at the point \( i \) and a bond vector \( s \) along the chain \( \hat{\textbf{b}}_{i+s} \). This dot product yields an angle \( \cos(\theta) \) between the bonds, which decreases as we move along the contour \( s \) until it becomes completely uncorrelated at \( \cos(\theta) = 0 \). In the the freely rotating chain model, this decay follows a decreasing exponential pattern, given by \( \langle\hat{\textbf{b}}_{i+s}\cdot\hat{\textbf{b}}_{i}\rangle = e^{-s/\ell_p} \), where \( \ell_p \) represents the persistence length. The averaging \( \langle...\rangle \) involves both a time and a sample average, incorporating different frames of the same worm, as well as averaging over trajectories of different worms.

To extract the persistence length from the bond-bond correlation function, we fit the exponential function to the data for the first 5 mm of the worm contour. Long-range correlations are less accurate because of fewer frames in which the entire worm is visible, often caused by overlaps or obstructing pillars and also deviations from ideal random walk model. An alternative method is to use the crossing point of \( \langle\hat{\textbf{b}}_{i+s}\cdot\hat{\textbf{b}}_{i}\rangle \) at \( 1/e \), but for very stiff worms, this crossing point may not always be reached for high contour lengths. By focusing solely on the first 5 mm of the contour, this issue is mitigated, and the data are less noisy at these data points. In figure S\ref{fig:PersistenceLength} the bond-bond correlation curves are reported.

\subsubsection{Effect of the geometry on $\tau_e$ and $R_e$}

\begin{figure}[h]
    \centering
    \includegraphics[width=\textwidth]{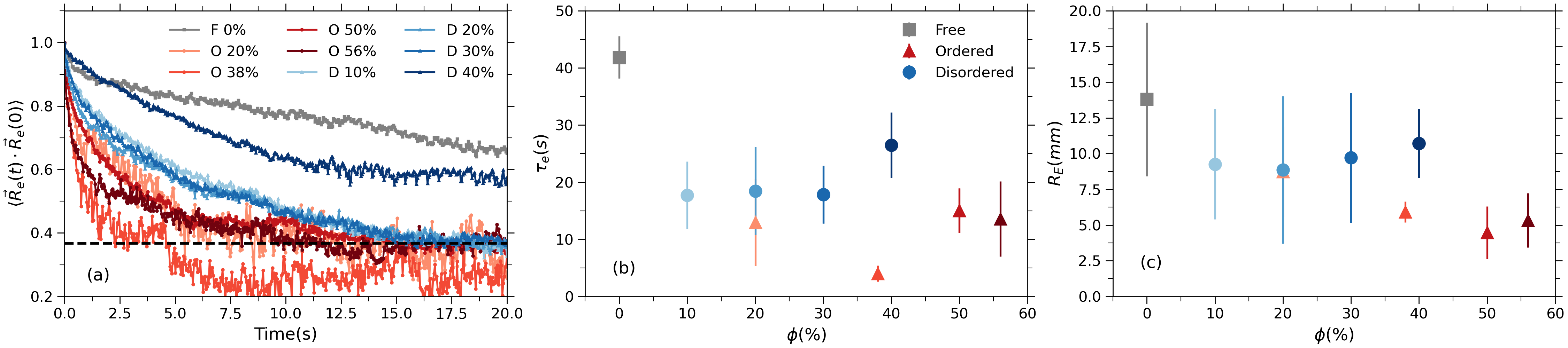}
    \caption{\textbf{Effect of the lattice geometry on $\tau_e$ and $R_e$.} (a) The orientational correlation in time, (b) the average reorientational decorrelation time, defined as the characteristic time at which the autocorrelation function of $R_e$ decays to $1/e$, and (c) the average end-to-end distance of the worms across all geometries. Error bars indicate the variance across the different measurements.}
    \label{fig:te_re}
\end{figure}

In order to apply the prediction for the long-time diffusion of tangentially driven chains \cite{Fazelzadeh2023}, we determined the average end-to-end distance $R_e$ and the reorientational decorrelation time $\tau_e$, both calculated from the detection and tracking of the worm's contour as described above. 
In all of our experiments and simulations, $\tau_e$ was defined as the characteristic time at which the autocorrelation function of $R_e$ decays to $1/e$. Using the mean squared rotational displacement (MSRD) $\Delta \theta^2(t)$ to estimate $\tau_e$ yielded consistent results with the autocorrelation method at short timescales. However, at longer timescales, the MSRD tends to plateau, complicating the extraction of a reliable $\tau_e$. This plateauing behavior likely arises from the quasi-2D nature of our worm system, where the orientation angle is not always well-defined. Instead, the $1/e$ cutoff in the time autocorrelation function consistently provides a robust and reliable timescale across all experimental conditions, that we use in the following. 
Sup.~Figs.~S\ref{fig:te_re}(a)-(c) show the orientational correlation over time, the average reorientational decorrelation time $\tau_e$, and the average end-to-end distance $R_e$ of the worms across all geometries, respectively. Notably, we observe no significant dependence of $\tau_e$ on the pillar surface fraction $\phi$, and only a weak dependence of $R_e$. This result may initially appear surprising, especially given the strong dependence of the worm's persistence length on $\phi$ (as discussed in Section C1 above and shown in Fig.~2(a) of the main text). However, this discrepancy can likely be attributed to challenges in accurately measuring $R_e$ and $\tau_e$ when portions of the worm are obscured by the pillars. Unlike tracking the center of mass or determining the persistence length, which remain well-defined even under partial obstruction, identifying the endpoints of the worm's skeleton for $R_e$ and $\tau_e$ measurements becomes challenging when they are hidden by the pillars.

\clearpage
\newpage
\subsection{Locomotion}

\subsubsection{Run lengths and trapping times}
\begin{figure}[!h]
    \centering
    \includegraphics[width=0.8\textwidth]{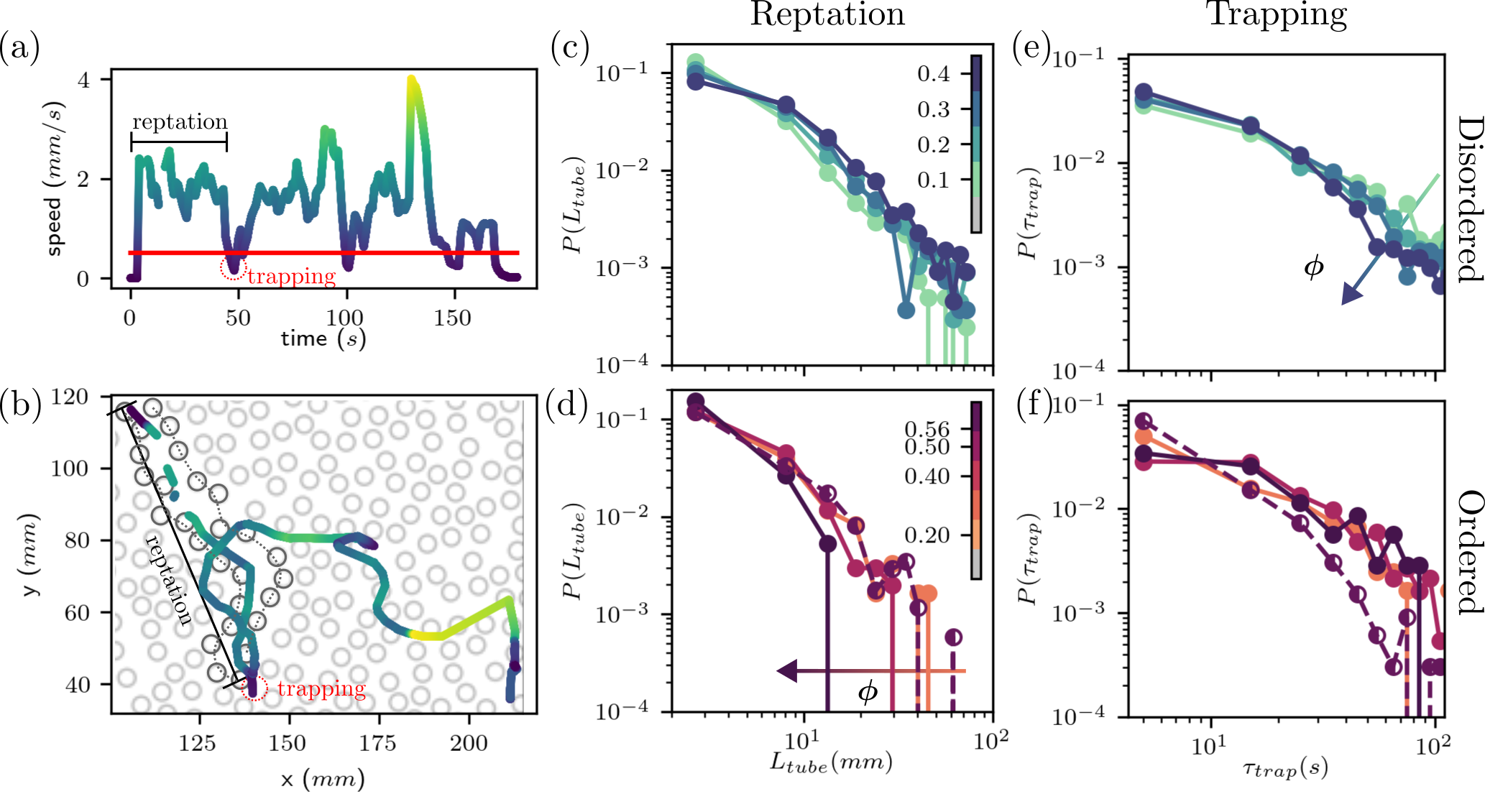}
    \caption{\textbf{Tube length and trapping time distributions} (a) Worms are defined as caged when their instantaneous speed drops below a cutoff of $0.5~mm/s$. Between caging events, worms crawl effective tubes of length $L_{tube}$. (b) Example of a worm trajectory ($\phi=40\%$) as it reptates within an effective tube made by the disordered positioning of pillars.  (c) and (d) Distribution of tube lengths for disordered (c) and ordered (d) media. Longer tube lengths are observed as $\phi$ increases in the disordered case, while the maximum tube length decreases for higher $\phi$ in the ordered medium. The opposite trend is observed for the distribution of trapping times in the disordered (e), and ordered media (f). The half-open symbols correspond to the ordered lattice at $\phi=50\%$.}
    \label{fig:runlength_50}
\end{figure}

In the main text, we excluded data for the ordered lattice at $50\%$ due to slight deviations from the trend. This deviation arises from the non-monotonous pore size distribution in a slightly noisy hexagonal lattice with a few removed pillars (see Sup.~Figs.~S\ref{SupFig:mazes} \& S\ref{SupFig:cavity_dist}). The worms exhibit a hopping-trapping behavior, hopping through short tubes between pores while occasionally finding longer tubes, which coincide with longer trapping times. Figure \ref{fig:runlength_50} reports the run length and trapping time values for this data.

\subsubsection{Effect of temperature}
\begin{figure}
    \centering
    \includegraphics{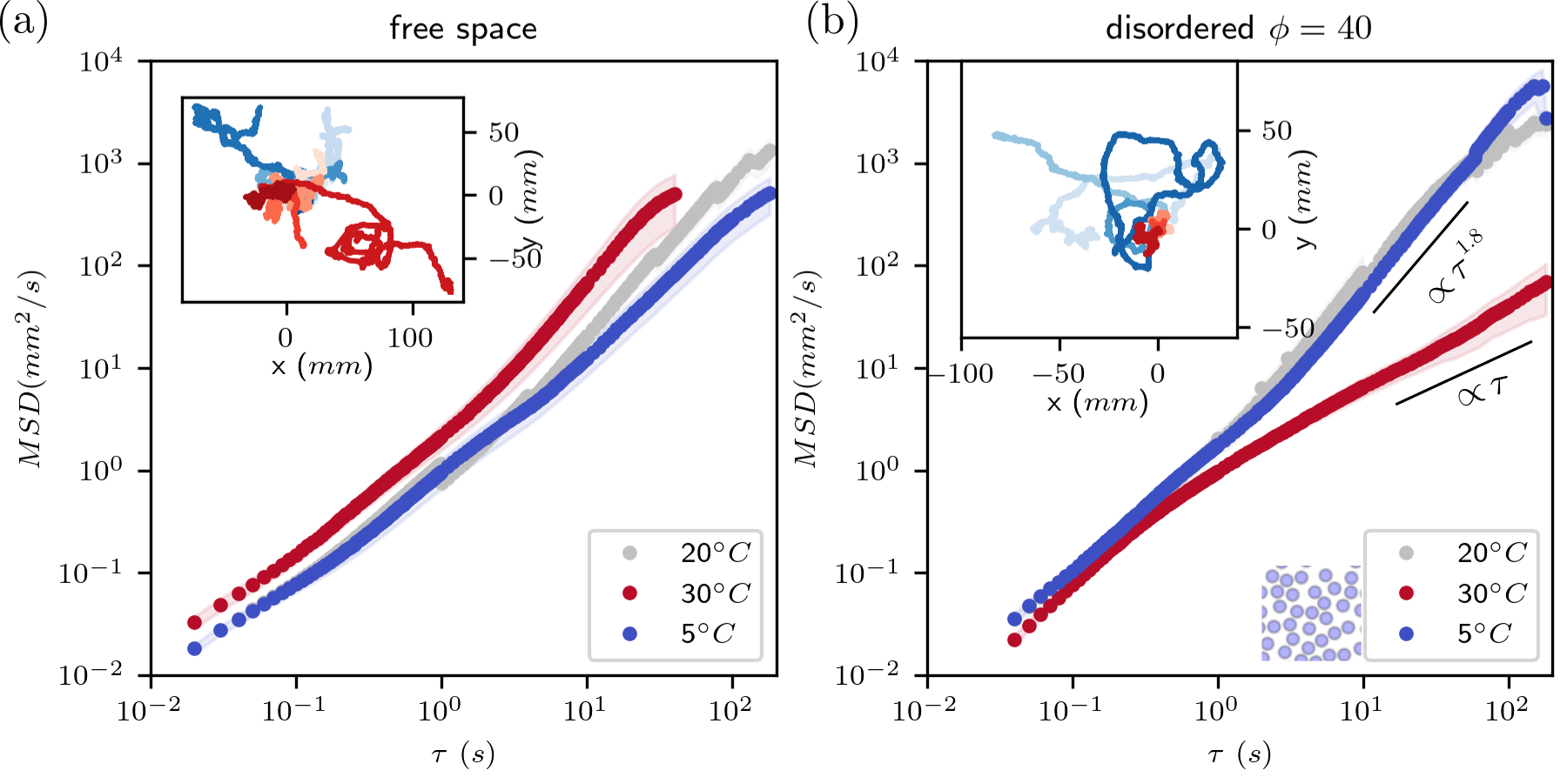}
    \caption{\textbf{MSD for the same set of worms at different temperatures in lab units.} a) Worms in free space. The hotter the worms, the more active they are and the higher their center of mass MSD-curves b) Worms in $\phi =40\%$ disordered porous media. The cold worms show strong reptation (even stronger than worms at room temperature), while the worms in a $30\celsius$ bath never reptate and show purely diffusive hopping-trapping behavior.}
    \label{fig:T_lab}
\end{figure}

\begin{figure}
    \centering
    \includegraphics[width=0.9\textwidth]{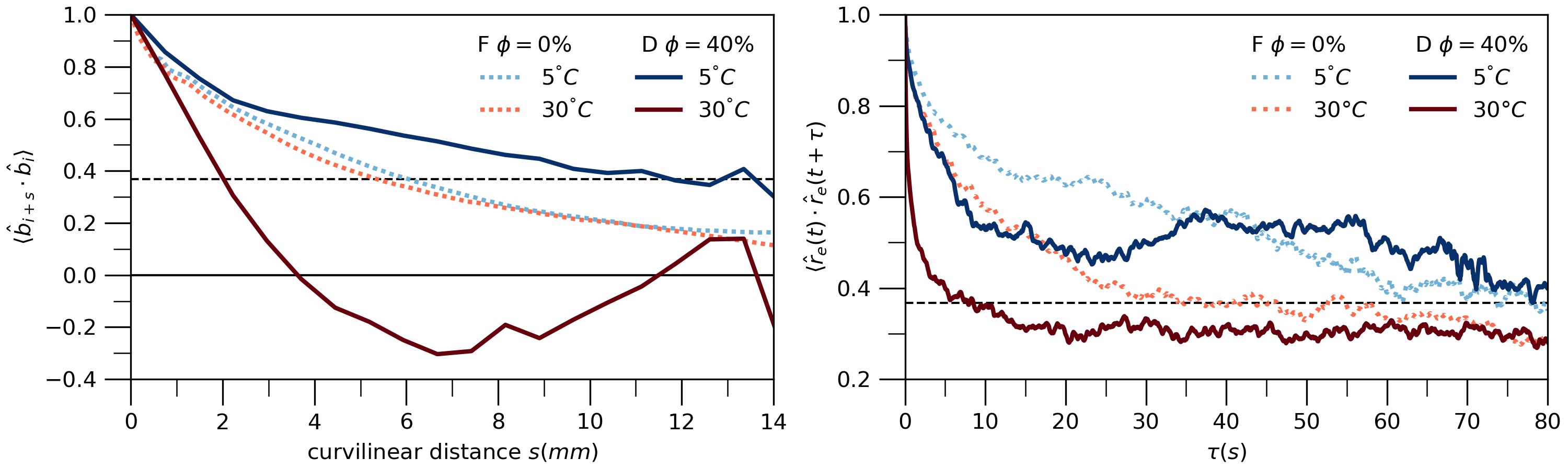}
    \caption{\textbf{Dependence of $\ell_p$ and $\tau_e$ on the temperature} a) bond-bond correlation for worms at $5\celsius$ (blue) and $30\celsius$ (red) as a function of time in free space (dotted lines) and the disordered media at $\phi$=40\% (solid lines). The intersection with the black dashed line at $1/e$ defines the persistence length. b) Orientational correlation for worms at $5\celsius$ (blue) and $30\celsius$ (red) as a function of time in free space (dotted lines) and the disordered media at $\phi$=40\% (solid lines). The intersection with the black dashed line at $1/e$ defines the reorientational relaxation time.}
    \label{fig:Tau_e_T}
\end{figure}

When placed in warmer water, our living worms become more active, resulting in faster motion and increased shape's fluctuation rates. In free space, their long-time diffusion increases with temperature \cite{Deblais2020a}. However, in dense disordered media, both the long-time diffusion time and the reorientational relaxation time decrease at higher temperatures. This occurs because the worms become trapped in cavities more frequently, preventing them from being stretched out long enough to initiate reptation. This trend is also evident in the effective persistence length of the worms. See~Sup.~Fig.~S\ref{fig:T_lab} for the MSD curves in laboratory units, Sup.~Fig.~S\ref{fig:Tau_e_T}(a) for the persistence length and Sup.~Fig.~S\ref{fig:Tau_e_T}(b) for the calculation of the reorientational relaxation time.

\clearpage
\newpage
\subsubsection{Bimodal distributions in the disordered geometries}
\begin{figure}[H]
    \centering
    \includegraphics[width=0.7\linewidth]{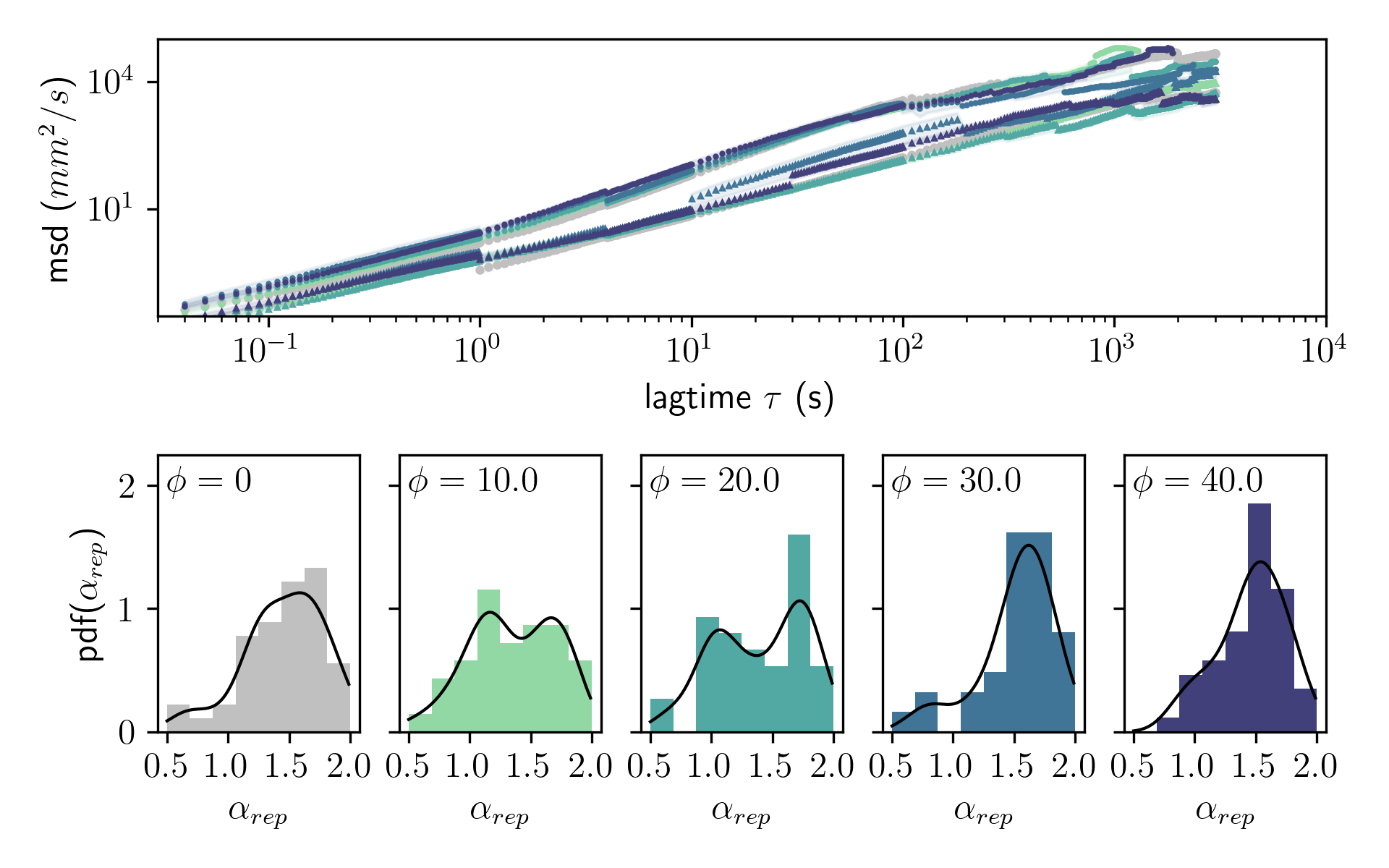}
    \caption{\textbf{Average MSD curves and slope distribution for the disordered media} For the disordered maze, some worms travel in large ballistic stretches, while some worms move along a diffusive trajectory. The probability of a worm moving ballistically increases if the worm is in a more crowded  environment (e.g., higher $\phi$).}
    \label{fig:MSD_populations}
\end{figure}

Here we would like to zoom in a bit more on the trajectories and subsequent MSD curves of the worms in the disordered geometries. Interestingly, distinct trajectories emerge in our experiments, segregating into two populations across all obstacle densities. One population showcases elongated, ballistic stretches as the worms reptate from one tunnel to another. In contrast, the second population exhibits hopping behavior, crawling from cavity to cavity with purely diffusive dynamics. As the obstacle density increases, the likelihood that a worm belongs to the ballistic population also increases. Consequently, the long-time diffusion constant increases when averaging over all trajectories from both populations, as illustrated in the Sup.~Fig.~S\ref{fig:MSD_populations}. In the top panel of the figure, for each maze configuration, the trajectories are segregated into two populations using a cutoff at the intermediate slope of each trajectory. It is evident that the worms exhibit ballistic and diffusive motion within each maze. However, as depicted in the PDFs in the bottom panels, there is a notable shift in behavior, with worms transitioning from a preference for diffusive motion to a preference for ballistic movement as obstacle density increases. This shift is attributed to the worm's capacity to reptate through effective tubes mapped out by the position of the pillars. It is worth noting here that the same individual worms were tested across all mazes, indicating behavioral changes in individual worms.

\clearpage
\section{SIMULATIONS}

\subsection{Phantom tangentially-driven polymer model}

\begin{figure}[H]
    \centering
    \includegraphics[width =0.7\linewidth]{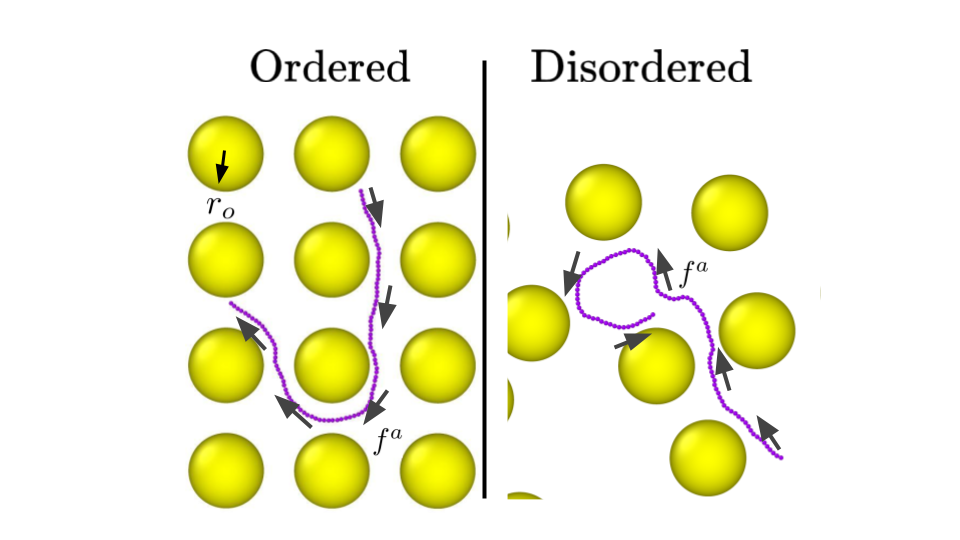}
    \caption{Schematic of the active tangentially-driven polymer in the ordered and disordered arrangement of obstacles.}
    \label{SupFig:Schematic}
\end{figure}

We implement the tangentially-driven polymer model~\cite{IseleHolder2015} into both a 2D ordered and disordered arrangement of circular obstacles, as illustrated in Sup.~Fig.S~\ref{SupFig:Schematic}. In our experimental setup, we observe the 2D projection of 3D active filaments around cylindrical pillars, indicating that the polymer can intersect with itself. To take into account this behavior, we neglect excluded volume interactions between monomers, and instead, we consider a phantom active polymer model comprising $N$ monomers. The motion of each monomer follows overdamped Langevin dynamics, described by:
  \begin{equation}
    \gamma \dot{\vec{r}}_i = - \sum_j \nabla_{\vec{r}_i} U+\vec{f}^a_{i}+\vec{f}^{r}_{i},
    \label{eq:brownian}
\end{equation}
where $\vec{r}_{i}$ is the position of the $i$th monomer, the dot denotes the derivative with respect to time and $\gamma$ is the friction coefficient between the bead and its surrounding medium. 

The potential energy $U$ of each monomer includes three different contributions. The 
first one is the harmonic spring potential $U_{\text{harmonic}}(r)=(k_s/2)(r-\ell)^2$, with equilibrium length $\ell$ and spring stiffness $k_s$ between adjacent monomers. The second part is the bending potential between each two neighboring bonds $U_{\text{bend}}(\theta_i)=\kappa(1-cos\theta_i)$, where $\theta_i$ denotes the angle between two consequent bonds intersecting at bead $i$ defined as $\theta_i= \cos^{-1} (\widehat{{t}}_{i,i+1} \cdot \widehat{{t}}_{i-1,i})$ with $\widehat{{t}}_{i,i+1}= \vec{r}_{i,i+1}/|\vec{r}_{i,i+1}|$ and $\vec{r}_{i,i+1}=\vec{r}_{i+1}-\vec{r}_{i}$.  Here, $\kappa$ is the bending stiffness and determines the intrinsic degree of flexibility of a polymer. Finally, the third contribution accounts for the excluded volume interactions between each bead and its surrounding obstacles. They are modeled by the short-ranged Weeks-Chandler-Andersen (WCA) potential~\cite{WCA}:
  \begin{equation}
 U_{\text{excl}}(r)=4  \epsilon \left[  (\frac{\sigma/2+r_o}{r})^{12} -(\frac{\sigma/2+r_o}{r})^6+\frac{1}{4}\right]
 \end{equation}
 for  $r< r_c=2^{1/6} (\sigma/2+r_{o})$,
where $\epsilon$ is the strength of the potential and has unit of energy, $\sigma$ is the diameter of the beads and $r_{o}$ is the radius of obstacles. The WCA potential is zero for interaction distances larger than the cutoff length $r_c$.\\

The active force on each bead, except for the end monomers, is given by: $\vec{f}^a_i=\frac{f^a}{2 \ell } (\vec{r}_{i-1,i}+\vec{r}_{i,i+1})$. The active force on the tail monomer is given by $\vec{f}^a_1=\frac{f^a}{2\ell } \vec{r}_{1,2}$  and for the head monomer by $\vec{f}^a_N=\frac{f^a}{2\ell } \vec{r}_{N-1,N}$. The random force is chosen as a white noise of zero mean and has the correlation $\langle \vec{f}^r_i(t) \cdot \vec{f}^r_j(t') \rangle=4 D_0 \gamma^2 \delta_{ij} \delta(t-t')$, where $D_0$ denotes the strength of noise of biological origin. It should be noted that the persistence length of a 2D passive ideal polymer in free space can be determined in terms of its bending stiffness and the strength of random force correlation as $\ell_p^0=2 \kappa \sigma/D_0 \gamma$.~\cite{Binder2020}.

We use the coordinates of the pillars in experiment to position the obstacles in a 2D simulation box with periodic boundary condition. We choose $l_u=\sigma$, $E_u=\epsilon$ and $\tau_u=\gamma \sigma^2/\epsilon$ with $\gamma=1$ as the units of length, energy, and time. Subsequently, we fix $\ell=1\sigma$, $N=100$, $r_o=8.33\sigma$ and the diffusion coefficient $D_0=1\epsilon/\gamma$. The ratio between the obstacle radius, chain length and monomer diameter is set with respect to the average length and thickness of the worms. We choose an active force of $f^a=0.1\epsilon/\sigma$ and the spring constants are chosen very stiff $k_s = 5000 \epsilon/\sigma \gg f^a/\ell$, to ensure that the mean bond length and the polymer contour length remain almost constant during simulations. It has been well established that the relaxation time of flexible tangentially driven chains in free space scales as $\tau_e\sim1/f^a$, while their enhanced diffusion coefficient scales as $D_l\sim f^a$~\cite{Fazelzadeh2023,Bianco2018}. Hence, when using the relaxation time as the unit of time, the $D_l$ of chains with different activities become identical. Our choice of active force ($f^a=0.1$), ensures that the activity dominates the motion of the whole polymer chain, but it is weak enough to let the random fluctuations affect the chain on monomer level. At the chain level, the thermal relaxation time suggested by the Rouse model is $\tau^{\text{Rouse}}\sim N^2=10^4$~\cite{doi1986theory}, which is $10$ times slower than the active relaxation time $\tau^{\text{Active}}\sim N/f^a=10^3$.
It is worth mentioning that any active force of the same order of magnitude would qualitatively give the same results as those given by $f^a=0.1$.

\subsection{MSD curves and input bending stiffness}

To gain insights from the tangentially driven polymer model, we kept certain input parameters fixed. In our simulations, spatial dimensions were determined relative to the setup size and the average length and thickness of the worms, as detailed earlier. Activity was intentionally set to low values to ensure that the contour fluctuations of the active polymer mirrored the fluctuations observed in the worms. Subsequently, the bending stiffness was derived from the persistence length of the worms (as described above). 
However, due to activity and interactions with obstacles (e.g. confinement), the effective persistence length, and thus the bending stiffness of the tangentially driven polymer, deviated from the input value. The bending stiffness was adjusted to ensure that the effective persistence length matched between experiments and simulations. The values are reported below:

\begin{table}[H]
\centering
\begin{tabular}{l|lll}
\label{tab:kappa}
Setup                 & $\kappa_{input}$ & $\kappa_{effective}$ & $\kappa_{experiments}$ \\ \hline
Free space, $\phi=0$  & 9                & 9.0                  & 9,17                   \\
Disordered, $\phi=10$ & 10               & 9.5                  & 9.36                   \\
Disordered, $\phi=20$ & 10               & 9.5                  & 9.31                   \\
Disordered, $\phi=30$ & 11               & 10.5                 & 10.95                  \\
Disordered, $\phi=40$ & 13               & 12.0                 & 11.95                  \\
Ordered, $\phi=40$    & 7                & 7.0                  & 7.17                   \\
Ordered, $\phi=50$    & 5.5              & 6.0                  & 6.20                  
\end{tabular}
\end{table}

\begin{figure}[H]
    \centering
    \includegraphics[width=\textwidth]{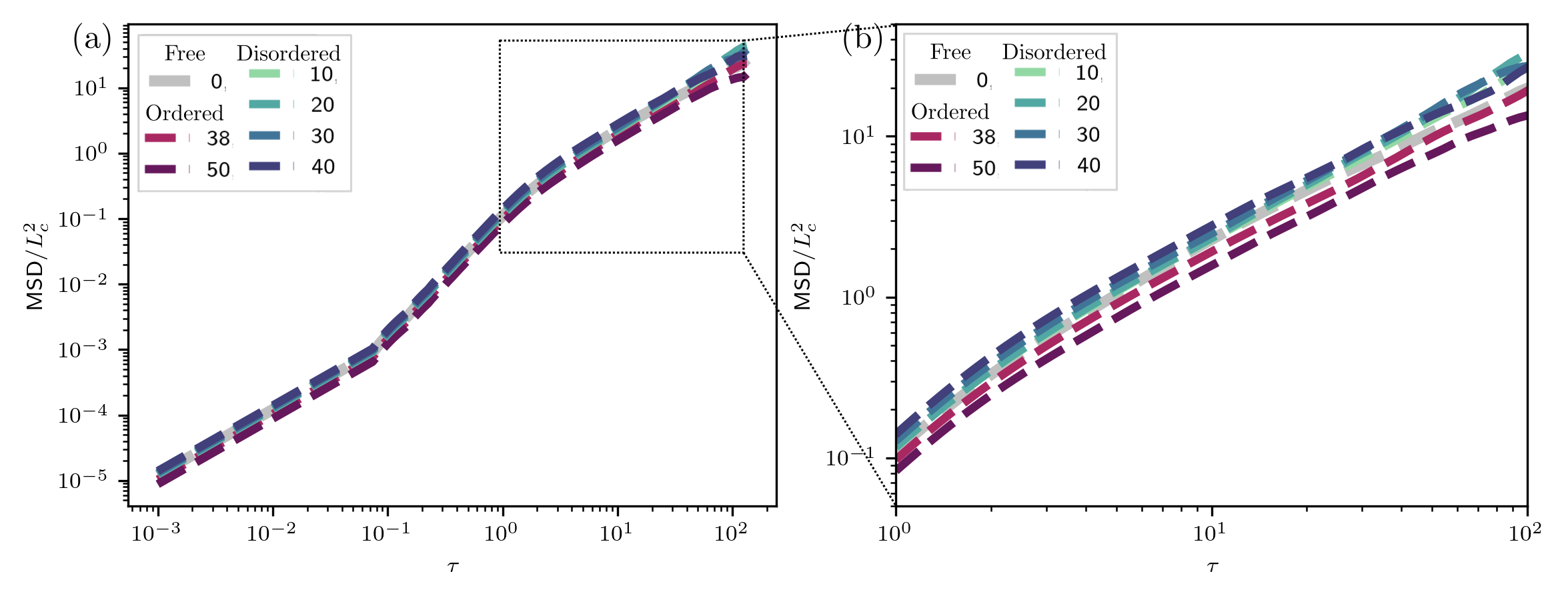}
    \caption{\textbf{MSD results from the simulations.} (a) All simulations are done in the same geometries as the experiments, with the relevant bending stiffness and rescaled by their respective rotational decorrelation time. See table \ref{tab:kappa} for all bending stiffness used. (b) Zoom to the diffusive part of the MSD.}
    \label{fig:msd_sim}
\end{figure}

In the Sup.~Fig.~S\ref{fig:msd_sim} the results from the simulations are shown, all curves are rescaled by their respective rotational decorrelation time. 

\subsection{Characterization of trapping events for simulation}
\begin{figure}
    \centering
    \includegraphics[width=0.6\textwidth]{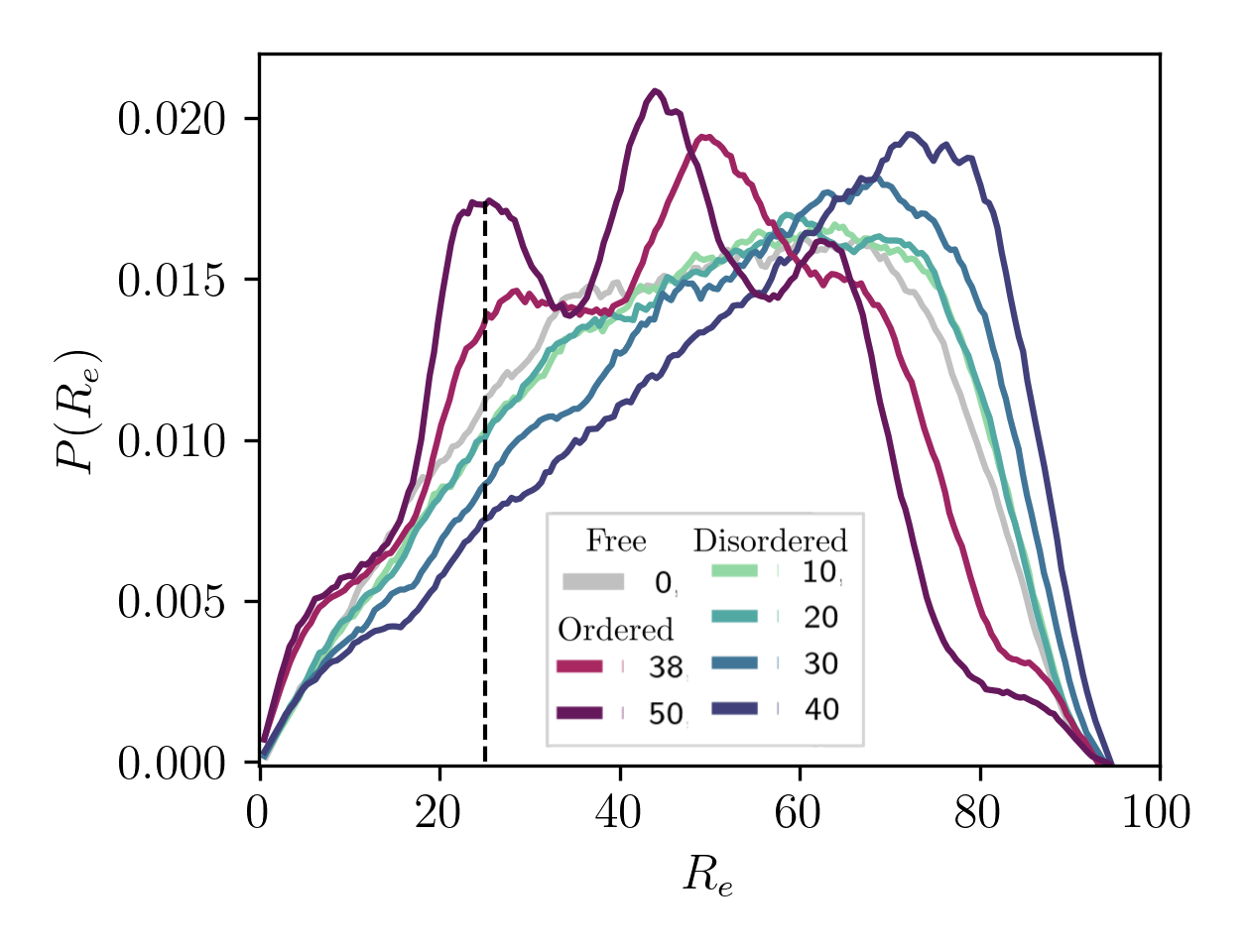}
    \caption{\textbf{The probability distribution function of end-to-end distance of simulated polymers in different media.} The black dashed line shows the value of the first peak ($R_e=25$) of the active polymer moving in ordered medium with $\phi=0.5$.}
    \label{fig:pdfRe}
\end{figure}

 We measure trapping and reptating events based on the following method for the simulated active polymers. Since for tangentially driven polymers the self propulsion velocity of the center of the mass is proportional to $R_e$, we use the distribution of the end-to-end distance to find the threshold on $R_e$ below which the chain is in trapped state. In Sup.~Fig.~S\ref{fig:pdfRe} we have the $P(R_e)$ for all of the simulations. The most confined chain is the one in the ordered medium with $\phi=0.5$. This particular distribution has three peaks showing the interaction of the polymer with the obstacles. We use the position of the first peak at $R_e\approx25$ to set the threshold. We use the same value for all of the simulations for consistency. Therefore, a chain is labeled trapped when $R_e<25$, otherwise reptating.

\begin{figure}
    \centering
    \includegraphics[width=\textwidth]{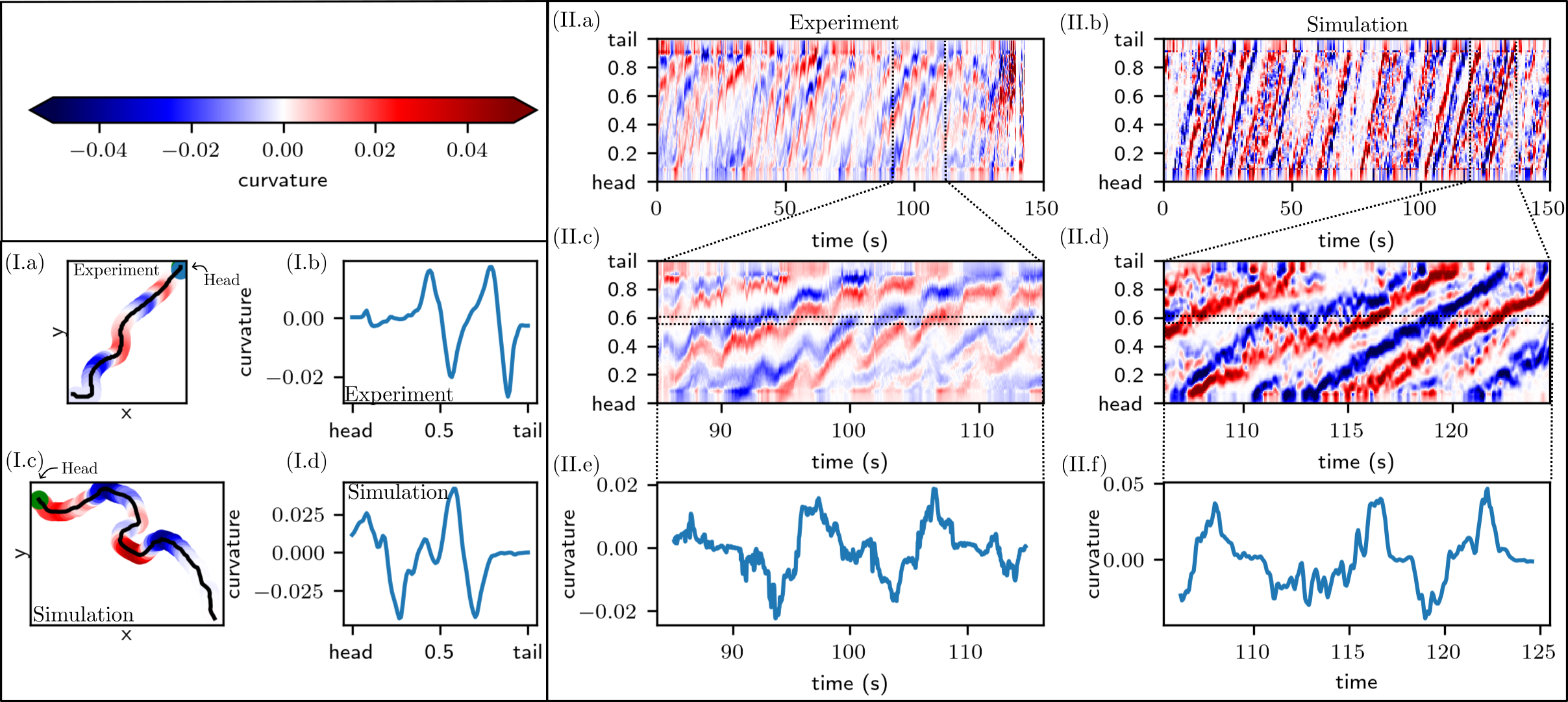}
    \caption{Principal component analysis (PCA) of the \textit{T.~tubifex} worm and the tangentially driven polymer. Panel (I) shows the extracted curvature profiles along the curvilinear length of the contour for the worm [(a), (b)] and the tangentially driven polymer [(c), (d)]. Panel (II) presents kymographs of the contour curvature over 150~s for the worm (a) and the tangentially driven polymer (b), with corresponding close-up views over a shorter timescale ($\sim$30~s) shown in (c) and (d). Panels (e) and (f) depict the curvature at a fixed segment located at a normalized curvilinear length of 0.6 from the head, for the worm and the polymer, respectively. Both systems exhibit multiple modes of curvature dynamics, with the worm revealing more complex bending behaviors as shown by the sawtooth like pattern in the curvature profile. (Simulation parameters: $\kappa$ = 7, $f^a$ = 0.1)).}
    \label{fig:kymo}
\end{figure}

\subsection{Principal component analysis of the \textit{T.~tubifex} worm and the tangentially driven polymer.}
From the comparison of the MSD measured in both simulations and experiments, we concluded that while the coarse-grained, long-time dynamics are very similar, significant differences emerge at shorter timescales. Here, we aim to investigate the origin of these differences. 
To better capture the detailed dynamics of both the living worms and simulated filaments, we analyzed the curvature of their backbones and tracked its evolution over time in the similar way as in \cite{Saggiorato2017}. In Sup.~Fig.~S\ref{fig:kymo}(I), we present snapshots showing the curvature profiles of the worm and the filament in free space. Sup.~Fig.~S\ref{fig:kymo}(II) illustrates the temporal evolution of curvature: for both systems, we observe waves propagating from head to tail at comparable speeds. To facilitate the comparison, the time in the simulations is rescaled to seconds using the reorientation timescale $\tau_e$ as previously defined. 
While the curvature waves travel nearly linearly for the simulated filaments, the worms exhibit a distinct second-order motion, indicating the presence of more complex bending modes. This highlights the need for a closer examination of active polymer models, potentially incorporating different orientations of active forces along the contour to account for transverse motions. Such refinements could better capture the richness of the observed dynamics.

\begin{figure}
    \centering
    \includegraphics[width=\textwidth]{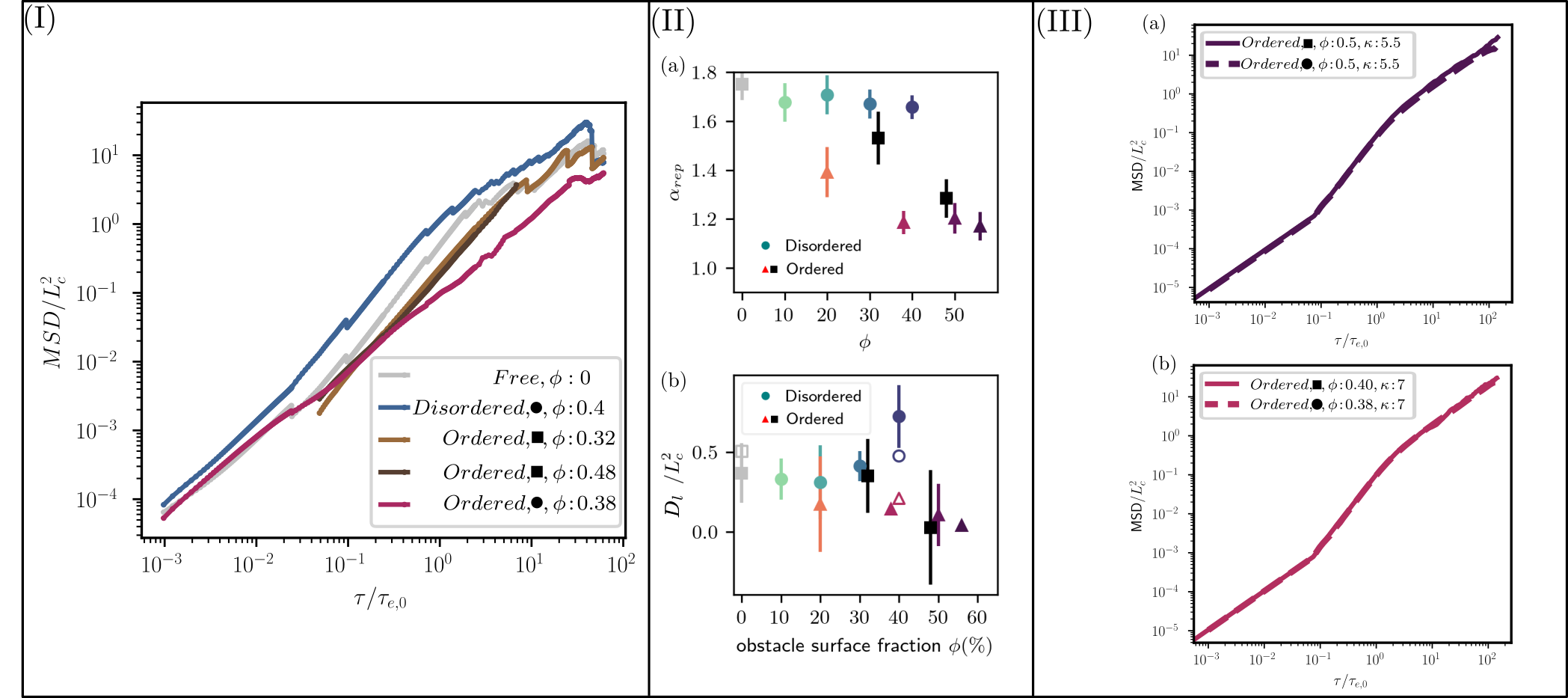}
    \caption{Effect of square pillars on the locomotion of worms [Panels (I), (II)] and tangentially driven polymers (III). Within the range of surface fractions considered, no significant differences are observed compared to the ordered circular pillar lattice discussed in the main text.}
    \label{fig:SquarePillars}
\end{figure}

\section{Effect of pillar shape on diffusion of active polymers in periodic lattices}
We conducted additional experiments using square-shaped pillars arranged in ordered lattices at surface fractions $\phi = 0.32$ and $\phi = 0.48$, with pillar dimensions of 2.5$\times$2.5 mm, following the same methods used for the  cylindrical obstacles. Complementary simulations of tangentially driven polymers under identical conditions were also performed. Results from both experiments and simulations indicate that worm dynamics in pillar lattices with square cross-sections are qualitatively similar to those in circular pillar lattices. Worms alternated between trapping in voids and reptating along confined channels, spending comparable amounts of time in each mode, as observed in the ordered arrays of cylindrical pillars (see Sup.~Fig.~S\ref{fig:SquarePillars}). The long-time diffusion coefficients for pillars with square and circular cross sections were consistent, showing no significant dependence on pillar shape for the surface fractions tested.

\newpage
\bibliography{refs}